\documentclass[letterpaper,twocolumn,10pt]{article}
\usepackage{usenix-2020-09}

\usepackage{algorithmic}
\usepackage{algorithm}
\usepackage{comment}
\usepackage{times}
\usepackage{amssymb}
\usepackage{graphicx}
\usepackage{longfbox}
\usepackage{fancyhdr}

\usepackage[font=small, textfont=bf]{caption}

\usepackage{amsmath}

\usepackage{makecell}

\usepackage{subcaption}

\usepackage{listings}

\usepackage{enumitem}
\usepackage{algorithmic}
\usepackage{booktabs} 
\usepackage{xspace}

\usepackage{microtype}
\usepackage[T1]{fontenc}
\usepackage[scaled]{beramono}

\usepackage{xcolor}
\usepackage{color}

\newcommand\unmarkedfootnote[1]{%
  \begingroup
  \renewcommand\thefootnote{}\footnote{#1}%
  \addtocounter{footnote}{-1}%
  \endgroup
}

\newcommand{\R}{\mathbb{R}}

\usepackage{lmodern}
\newcommand{\tc}[1]  {{\ttfamily\fontseries{m}\selectfont #1}}
\newcommand{\tbf}[1]  {{\bf #1}}

\newcommand{\parheader}[1] {\noindent\textbf{#1}}

\definecolor{mred}{rgb}{.80,.12,.30}
\definecolor{grey}{rgb}{0.5,0.5,0.5}
\definecolor{purple2}{rgb}{.75,0,.85}
\definecolor{pistachio}{rgb}{0.58, 0.77, 0.45}
\definecolor{steelblue}{rgb}{.20,.35,.90}
\newcommand{\suman}[1]{\textcolor{mred}{(Suman: #1)}}
\newcommand{\gabe}[1] {\textcolor{purple2}{(Gabe: #1)}}
\newcommand{\abhi}[1] {\textcolor{blue}{(Abhi: #1)}}
\newcommand{\ksb}[1]  {\textcolor{pistachio}{(Koustubha: #1)}}
\newcommand{\dongdong}[1]  {\textcolor{pistachio}{(Dongdong: #1)}}

\renewcommand{\suman}[1]{}
\renewcommand{\gabe}[1]{}
\renewcommand{\abhi}[1]{}
\renewcommand{\ksb}[1]{}
\renewcommand{\dongdong}[1]{}

\newcommand{\trm}[1]{\textrm{#1}}
\newcommand{\mbf}[1]{\mathbf{#1}}
\newcommand{\prox}{\operatorname{prox}}
\newcommand{\op}{\operatorname{op}}

\newcommand{\revise}[1] {#1}
\newcommand{\rrevise}[1] {#1}
\newcommand{\rv}[1] {#1}
\newcommand{\rvv}[1] {#1}

\newcommand\LSTSmall{\fontsize{8}{8.2}\selectfont}
\newcommand*\MyLSTfont{\LSTSmall\ttfamily\SetTracking{encoding=*}{-60}\lsstyle}

\definecolor{codegreen}{rgb}{0,0.9,0}
\definecolor{codered}{rgb}{0.9,0.0,0}
\definecolor{codeblue}{rgb}{0,0,0.8}
\definecolor{codegray}{rgb}{0.5,0.5,0.5}
\definecolor{codepurple}{rgb}{0.58,0,0.82}
\definecolor{backcolour}{rgb}{0.95,0.95,0.92}

\lstdefinestyle{braystyle}{
  commentstyle=\color{codered},
  keywordstyle=\color{codeblue},
  numberstyle=\color{black},
  stringstyle=\color{codepurple},
  frame=single,
  numbersep=5pt,
  keepspaces=true,        
  numbers=left,
  language=C,
  basicstyle=\MyLSTfont,
  xleftmargin=2em,
  framexleftmargin=2em,
}

\begin{document}

\date{}

\title{\Large 
Fine Grained Dataflow Tracking with Proximal Gradients }

\author{
{\rm Gabriel Ryan$^\dag$, Abhishek Shah$^\dag$, Dongdong She$^\dag$, Koustubha Bhat$^\ddag$, Suman Jana$^\dag$}\\
$^\dag$Columbia University, $^\ddag$Vrije Universiteit Amsterdam
} 

\maketitle
\begin{abstract}
Dataflow tracking with Dynamic Taint Analysis (DTA) is an important method in systems security with many applications, including exploit analysis, guided fuzzing, and side-channel information leak detection. However, DTA is fundamentally limited by the Boolean nature of taint labels, which provide no information about the significance of detected dataflows and lead to false positives/negatives on complex real world programs.

We introduce proximal gradient analysis (PGA), a novel, theoretically grounded approach that can track more accurate and fine-grained dataflow information. 
PGA uses proximal gradients, a generalization of gradients for non-differentiable functions, to precisely compose gradients over non-differentiable operations
in programs. Composing gradients over programs eliminates many of the dataflow propagation errors that occur in DTA and provides richer information about how each measured dataflow effects a program.

We compare our prototype PGA implementation to three state of the art DTA implementations on 7 real-world programs. Our results show that PGA can improve the F1 accuracy of data flow tracking by up to $33\%$ over taint tracking \rv{(20\% on average)} without introducing any significant overhead ($<5\%$ on average). We further demonstrate the effectiveness of PGA by discovering 22 bugs (20 confirmed by developers) and 2 side-channel leaks, and identifying exploitable dataflows in 19 existing CVEs in the tested programs. 
\unmarkedfootnote{To appear in USENIX Security 2021.}
\end{abstract}

\section{Introduction}
Dataflow analysis with dynamic taint analysis (DTA) is a fundamental building block in many common systems security tasks, such as automated vulnerability analysis, guided fuzzing, discovering information leaks, and malware analysis~\cite{clause2007dytan, schwartz2010all, ganesh2009taint, vuzzer, arzt2014flowdroid, yan2012droidscope}. DTA analyzes dataflow between a specified set of sources and sinks in a program by instrumenting the program and tracking taint as it executes~\cite{newsome2005dynamic,jflow}. 

However, DTA is fundamentally limited by the Boolean information contained in taint labels: data either is tainted by a given source or not; there are no intermediate states or other sources of information. This means there is no way to identify and prioritize which dataflows are most significant.
For example, given a series of operations \tc{x1 = a*8; x2 = b/8; y = x1 + x2;} changes to the value of \tc{a} will have a larger effect on the value of \tc{y} than changes to value of \tc{b}, but taint labels cannot make this distinction.
Moreover, it limits the ability of DTA frameworks to account for dataflows that are dependent on how operations compose. For example, in \tc{x1 = x * 2; x2 = x1 \& 1;}
variable \tc{x2} will only be affected by changes in the first bit of \tc{x1}, but changes to \tc{x} will not affect \tc{x2} due to the intermediate multiplication by 2. 

While most DTA systems incorporate some special rules to handle these types of cases, we find in our evaluation (Section \ref{sec:accuracy}) that current DTA systems with these rules still make many errors in predicting dataflows, even at high compiler optimization levels that eliminate most intermediate operations. 
 These errors have prevented DTA from being successfully applied in applications such as detecting keyloggers and memory corruption attacks ~\cite{balzarotti2008saner, slowinska2009pointless, slowinska2010pointer, taintinduce}.


\revise{The limitations of DTA 
led several researchers to propose Quantitative Information Flow (QIF) based methods as a more fine grained form of dataflow ~\cite{newsome2009measuring}.
However, while QIF is able to track data more precisely, computing these measures is computationally expensive and does not scale effectively to large programs ~\cite{kopf2010approximation}.}

In this paper, we propose an alternate measure of dataflow 
that addresses the limitations of DTA while retaining its advantages in scalability. We observe that {\bf gradient}, a multi-variate generalization of derivatives from elementary calculus, is a popular method for tracking the influence of inputs through differentiable models ~\cite{cook1986assessment}. \revise{In particular, gradients have been used in neural networks to perform a variety of tasks that are analogous to the applications of DTA in program analysis,} including generating inputs to trigger errors, explaining output behaviors, and maximizing test coverage ~\cite{ baehrens2010explain, simonyan2013deep, shrikumar2017learning, goodfellow2014explaining, pei2017deepxplore, tian2018deeptest}. 


\begin{figure}[t]
  \centering
  \includegraphics[width=\linewidth]{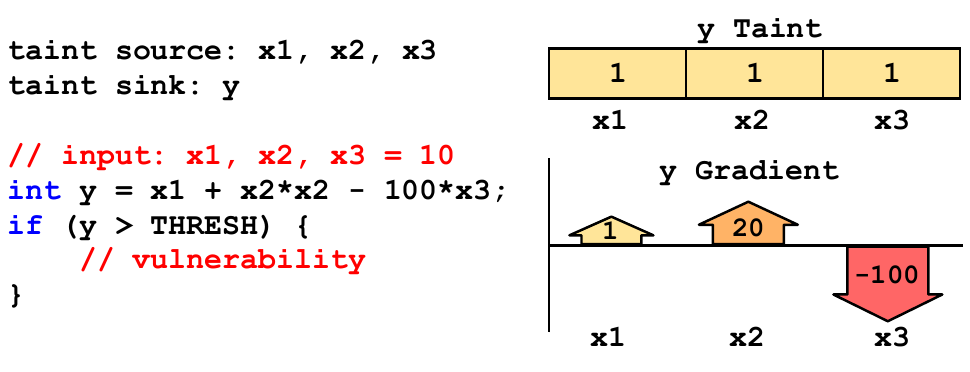}
  \vspace{-15pt}
  \caption{\label{fig:example1} \revise{Example program in which gradient can guide a search to reach a vulnerability. While taint tracking identifies \tc{y} as tainted by all three inputs, gradient measures the magnitude and direction of each influence, identifying that \tc{x3} is the most influential input and that minimizing it will maximize \tc{y} due to its negative gradient.}}
  \vspace{-15pt}
\end{figure}

The additional information provided by gradients confer two crucial advantages: (i) {\bf Fine-grained tracking.} \revise{Gradients measure both the {\it magnitude} and {\it direction} of influence, which indicate how changes to an operation's input will effect its output. This means gradients can be used to identify which marked sources are most influential, and how they will effect program behavior. This is illustrated in Figure \ref{fig:example1}, in which the magnitude of the gradient identifies the most influential input, and the direction of the gradient indicates how that input can be changed to reach a vulnerability.} (ii) {\bf Precise composition.} 
Gradients can be used to identify when an operation input will have no effect on its output due to composition.
\rv{For \tc{x1 = x * 2; x2 = x1 \& 1;} the gradient of \tc{x1} will be 2 and the gradient of \tc{x2} will be 0,} which correctly identifies that the first bit will never change in the operation \tc{x1 \& 1} and therefore there will be no dataflow.

\begin{figure*}[t]
  \centering
  \begin{subfigure}[b]{0.24\textwidth}
    \includegraphics[width=\textwidth]{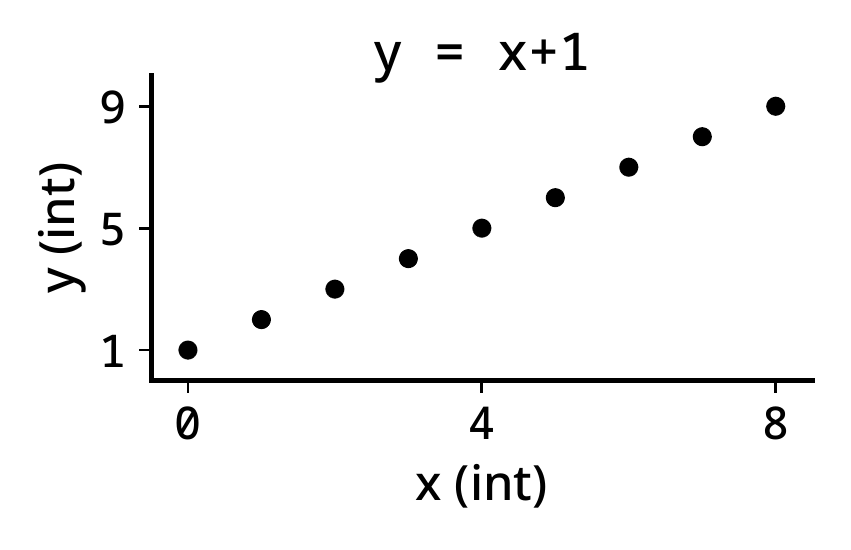}
  \end{subfigure}
  \begin{subfigure}[b]{0.24\linewidth}
    \includegraphics[width=\linewidth]{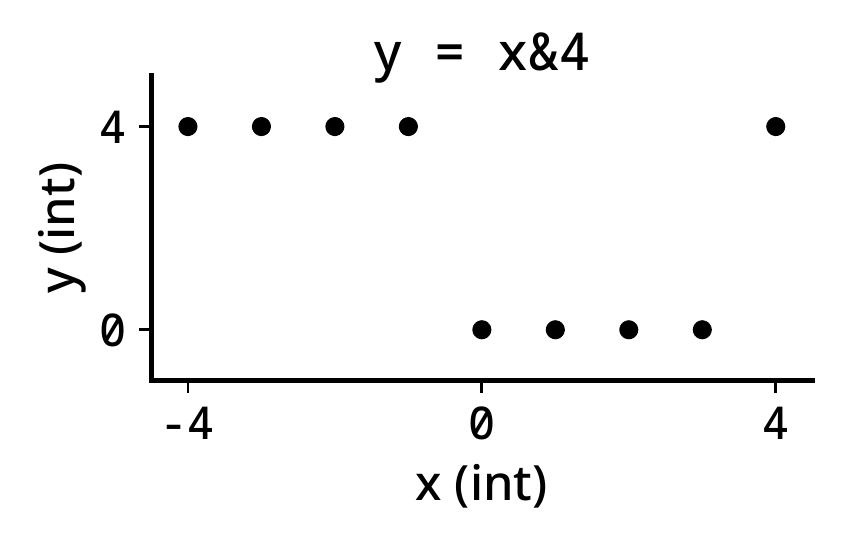}
  \end{subfigure}
  \begin{subfigure}[b]{0.24\linewidth}
    \includegraphics[width=\linewidth]{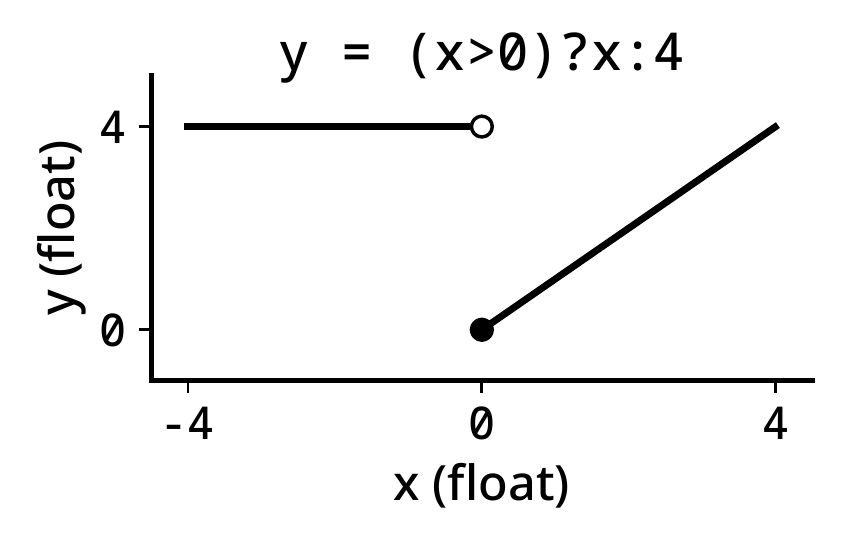}
  \end{subfigure}
  \begin{subfigure}[b]{0.24\linewidth}
    \includegraphics[width=\linewidth]{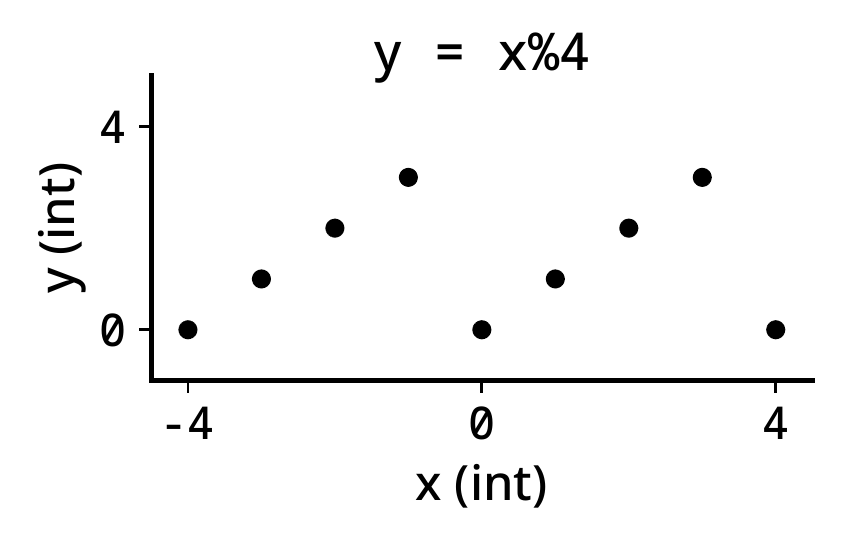}
  \end{subfigure}
  \vspace{-5pt}
  \caption{ Different types of discrete and discontinuous operations that occur in real-world programs}
  \label{fig:ex_funcs}
  \vspace{-10pt}
\end{figure*}

However, in general, programs contain many non-differentiable operations with different types of non-smooth behavior (e.g. bitwise operations, integer arithmetic, and branches as shown in Figure~\ref{fig:ex_funcs}) that cannot be differentiated directly.
Therefore, we build on the rich non-smooth calculus literature to define generalized gradients for programs 
that satisfy weaker forms of chain rule
~\cite{ward1991chain, griewank1995automatic, nesterov2005lexicographic}. 
To evaluate generalized gradients on programs, we use {\bf proximal gradients}, which
compute gradient on non-differentiable operations by finding the local minima ~\cite{parikh2014proximal}. 
Proximal gradients provide a theoretically grounded framework for gradient evaluation that allows us to precisely track dataflow across real-world programs with minimal compositional errors.

We implement a prototype of Proximal Gradient Analysis (PGA) as an LLVM pass that instruments programs during compilation to compute proximal gradients. We compare PGA to three state-of-the-art DTA systems on 7 widely used applications and show that PGA achieves up to 33\% better F1 accuracy \rv{(20\% on average)} than DataFlowSanitizer, the best performing DTA system, without incurring any significant (<5\%) extra overhead. We apply PGA to guided fuzzing and show that using PGA achieves up to 56\% higher edge coverage \rv{(10\% on average)} than DTA in a controlled comparison, as well as improving the coverage achieved by a state-of-the-art fuzzer \texttt{NEUZZ} by 13\% on average ~\cite{neuzz}. Finally, we use PGA to discover 22 bugs and 2 side-channel leaks, and analyze 19 existing CVEs.

The rest of this paper is organized as follows. First, Section~\ref{sec:background} summarizes the background on different generalizations of gradients to non-smooth analysis. Next, we describe our methodology for computing proximal gradients on real-world programs in Section~\ref{sec:methodology}. We describe the details of our implementation of proximal gradient analysis in Section~\ref{sec:implementation}, Section~\ref{sec:evaluation} contains the details of our evaluation setup and results, and we discuss the tradeoffs of PGA and DTA in Section \ref{sec:discussion}. Finally, we summarize related work in Section~\ref{sec:related} and conclude in Section~\ref{sec:conclusion}.

Our main contributions are:
\vspace{-3pt}
\begin{enumerate}
\item We are the first, to the best of our knowledge, to use non-smooth analysis for dataflow tracking in real-world programs. Specifically, we design, implement, and evaluate Proximal Gradient Analysis (PGA), a novel, theoretically grounded technique for \revise{measuring fine grained influence}  
in real-world programs.

\item We implement our PGA framework for automatically computing and tracking proximal gradients as an LLVM pass. \rv{An open source release of PGA is available at \url{https://github.com/gryan11/PGA}.} \abhi{not active? maybe say will be released at? or put placeholder?}

\item We perform extensive experimental evaluation of PGA and compare it to three state of the art DTA implementations, DataFlowSanitizer, \texttt{libdft}, and \texttt{Neutaint}, on 7 popular, real-world programs. PGA achieves up to 33\% higher F1 accuracy than DTA \rv{(20\% on average)} without introducing significant additional overhead (on average <5\%). PGA also achieves up to 56\% improvement in new edge coverage relative to DTA \rv{(10\% on average)} for data-flow-guided fuzzing, as well as improving the coverage achieved by a state-of-the-art fuzzer \texttt{NEUZZ} by 12.9\% on average.

\item We demonstrate that PGA's fine-grained tracking is helpful for finding and analyzing different types of bugs and information leaks. In our experiments, PGA found 22 bugs and 2 side-channel leaks in our tested programs. PGA also detected the exploitable dataflow in 19 known CVEs, including 2 where DTA fails.
\end{enumerate}
\vspace{-5pt}

\section{Background}
\label{sec:background}

Our approach to gradient-based dataflow analysis draws on several techniques from the mathematical analysis and optimization literature. We provide a summary of the relevant methods below. We first summarize standard methods for computing gradients over compositions of smooth functions, and then review techniques from the non-smooth analysis literature that can be applied to computing gradients over programs.


\subsection{Smooth Analysis}

\noindent {\bf Gradients.} 
The derivative for a smooth scalar function $f(x)$ is defined as $f'(x) = \mathop {\lim }\limits_{\delta x \to 0} \frac{{f( {x + \delta x }) - f\left( x \right)}}{\delta x}$, where $f: \mathbb{R} \rightarrow \mathbb{R}$. 
If a function has a derivative for all points in its domain, then it is considered a differentiable function. The gradient is a generalization of the derivative to multi-variate functions, where $f: \mathbb{R}^n \rightarrow \mathbb{R}$ and $\nabla f: \mathbb{R}^n \rightarrow \mathbb{R}^n$, that can be understood as the slope of the function at the point where it is evaluated. When a function is vector-valued (i.e. $f: \mathbb{R}^n \rightarrow \mathbb{R}^m$), the Jacobian generalizes gradient by evaluating the gradient of each of the $m$ outputs: $\mathcal{J} f: \mathbb{R}^n \rightarrow \mathbb{R}^{n\times m}$. For the rest of the paper, functions are multi-variate unless otherwise noted.  


\noindent {\bf Chain Rule.}
Gradients of compositions of differentiable functions can be computed from gradients of the individual functions. \abhi{maybe state that no "limits" are computed, b/c these gradients are known analytically? of these individual functions}
This is known as the chain rule of calculus and is defined as follows, where $\circ$ indicates the composition of two functions $f$ and $g$, and $f'$ and $g'$ are their respective gradients: 
\vspace{-2pt}
\begin{align}
  (f \circ g)' &= (f' \circ g) * g' \label{eq:chain_rule}
\end{align}
Elementwise multiplication is used when $f$ and $g$ are multivariate. \abhi{my understanding was that it was an inner product?}

\noindent \textbf{Automatic Differentiation.}
Automatic Differentiation (AutoDiff) uses the chain rule to compute the gradient for potentially large compositions of differentiable functions.
AutoDiff has been a longstanding tool in computational modeling and is a core component of deep learning frameworks such as Tensorflow ~\cite{Wengert:1964:SAD:355586.364791, tensorflow2015-whitepaper}. However, existing AutoDiff methods and frameworks are limited to working with mostly continuous functions with limited discontinuity (e.g. ReLUs in neural networks). \abhi{relu is continuous but not differentiable?}


\subsection{Non-smooth Analysis}
\label{sec:nonsmooth_background}

Extensive work has been done in the field of mathematical analysis on methods for approximating gradients over 
non-smooth functions. \abhi{maybe clarify non-differentiable implies non-smooth, ie smooth -> differentiable}
In this section we consider general multivariate functions of type $f:\R^n \rightarrow \R$.
We first describe a generalized type of continuity, called Lipschitz continuity, \abhi{i believe it's a specific type not a generalized type} that applies to non-smooth operations in programs, and then define a generalization of gradients that apply to Lipschitz continuous functions.

\begin{figure}
  \centering
  \begin{subfigure}[b]{0.45\columnwidth}
        \includegraphics[width=\linewidth]{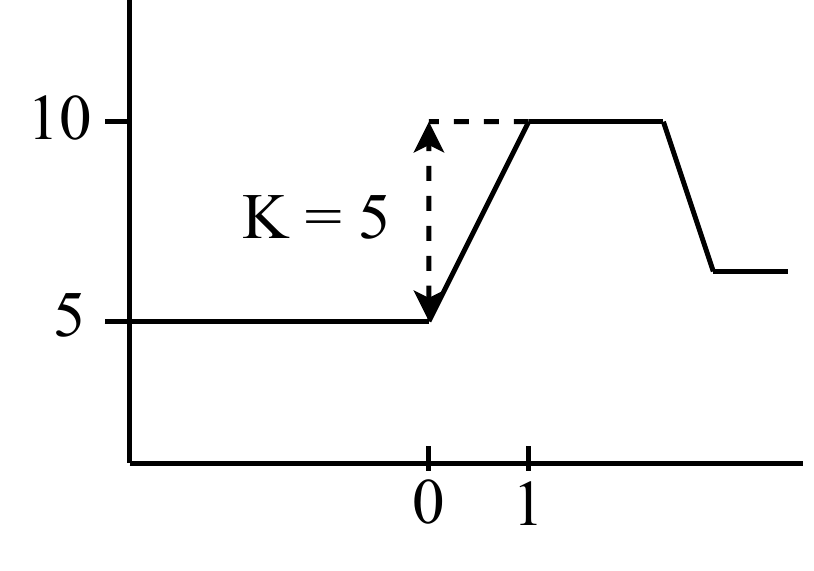}
    \caption{\label{fig:lipschitz_ex} Lipshitz func.}
  \end{subfigure}
  \hspace{0.10cm}
  \begin{subfigure}[b]{0.45\columnwidth}
        \includegraphics[width=\linewidth]{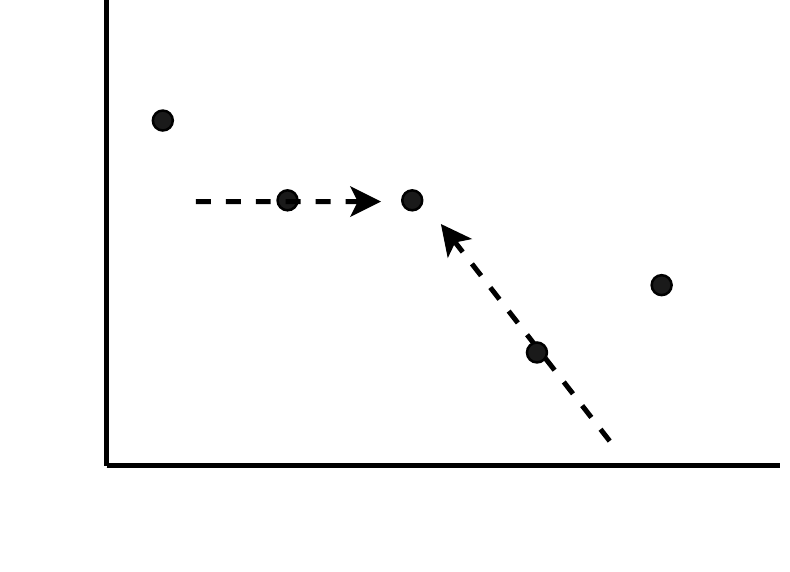}
  \caption{\label{fig:dir_deriv} Directional deriv.}
  \end{subfigure}
  \vspace{-1pt}
     \setlength{\belowcaptionskip}{-20pt}
  \caption{\label{fig:grad_exs} Example of a Lipschitz function with K=5 and directional derivatives on a discrete function.}
  \vspace{5pt}
\end{figure}

\noindent \textbf{Lipschitz Continuity.} A function is Lipschitz continuous if its output does not change too much for small changes in the input. 
 Formally, a function $f$ is Lipschitz continuous if there exists a constant $K$ (called the Lipschitz constant) that bounds how much the value of $f$ can change between any two points in its domain.
Figure~\ref{fig:lipschitz_ex} shows a simple Lipschitz continuous function along with the corresponding Lipschitz constant. 
In general the operations in any useful computation \abhi{maybe say "on real-world hardware"?} will yield a Lipschitz continuous function.  

\noindent \textbf{Generalized Gradients.}
On Lipschitz continuous functions, {\it generalized gradients} are used to approximate \abhi{consider changing word choice?} gradients~\cite{clarke1990optimization, rockafellar2009variational}. Generalized gradients consist \abhi{I would change to "can be thought of", b/c they do not consist of them} of generalized directional derivatives, 
which evaluate the gradient in a single direction as shown in Figure~\ref{fig:dir_deriv}. A generalized directional derivative in a direction $\mathbf{v} \in \R^n$ is defined as follows:
\vspace{-2pt}
\begin{align}
  f' \left(x; \mathbf{v}\right) = \lim \sup_{y \rightarrow x, \lambda \downarrow 0} \frac{f\left(y + \lambda \mathbf{v}\right) - f\left(y\right)}{\lambda} \label{eq:generalized_dir_deriv}
\end{align}
\noindent Here $x$ and $y$ are two points in the domain of $f$ where $x$ is the point the derivative is evaluated, and $\lambda$ is a distance along the vector $\mathbf{v}$ that the derivative is taken in. 
\abhi{i would say chain rule for a common type of generalized directional derivatives to be safe}
\rv{The chain rule for directional derivatives with functions $g:\R^n \rightarrow \R^n$ and $f:\R^n \rightarrow \R$ is defined:}
\begin{align}
    (f \circ g)' (x; \mathbf{v}) = f' \left(g(x); g' (x; \mathbf{v}) \right)
    \label{eq:directional_chain_rule}
\end{align}
When applied to generalized directional derivatives the composing functions must be monotonic. Several relaxed versions of the chain rule apply to generalized derivatives under different weaker assumptions about the composite functions \cite{ward1991chain,griewank1995automatic,nesterov2005lexicographic}.  \abhi{probably don't need this paragraph at all?}

\abhi{maybe mention this is due to definition vs computational matters, in theory no need to approximate}A generalized gradient is approximated with a set of directional derivatives based on a matrix $\mathbf{V} \in 
\R^{n\times p} = \left[\mathbf{v}_1, \mathbf{v}_2, \ldots, \mathbf{v}_p\right]$ of $p$ vectors in the domain of $f$ representing the directions in which the derivatives are evaluated.
\begin{align}
     f' (x; \mathbf{V}) = \left[f'(x; \mathbf{v}_1), f'(x;\mathbf{v}_2), \ldots,  f'(x;\mathbf{v}_p)\right]
        \label{eq:generalized_gradient}
\end{align}
When $f$ is a composition of functions, the chain rule from Eq.~\ref{eq:directional_chain_rule} can be applied to each of the generalized directional derivatives:
\begin{align}
     (f \circ g)' (x; \mathbf{V}) = [&(f \circ g)' (x; \mathbf{v}_1)),...,(f \circ g)' (x; \mathbf{v}_p)] \label{eq:directional_chain_rule_gradient}
\end{align}

\abhi{maybe state you use the word gradient but the images show the case when dimension = 1. Also, maybe state explicitly the gradient is a set of directional derivatives in this paper}

\section{Methodology}
\label{sec:methodology}

At a high level, our gradient propagation framework, PGA, is similar to Autodiff, computing the gradient of each operation and using the results as inputs to the next gradient computation. However, unlike Autodiff, we approximate the gradients of discrete \abhi{maybe make consistent with Intro/Background} functions with proximal gradients. 

\noindent \textbf{Proximal Gradients.} 
Since programs are generally composed of discrete operations on integers, 
we define a gradient approximation called proximal gradients that can be evaluated on these discrete functions. 
Proximal gradients use the minima of a function within a nearby region  \abhi{see above on the word discrete}
defined with a special operator called the proximal operator
~\cite{parikh2014proximal}. This can be evaluated on both discrete and continuous functions $f:X^n \rightarrow X$, where $X$ is a set with euclidean norm that can represent integers or floats. \abhi{nice, maybe capitalize Euclidean, there is no such thing as a discrete function from my readings? maybe define it since it is used a lot}
\begin{align}
\label{eq:prox_op}
  \prox_f\left(x\right) &= \arg\min_y \left(f\left(y\right) + \tfrac{1}{2}|| x-y||_2^2\right)
\end{align} 
The notation $\arg\min_y$ indicates that the operator evaluates to the value of $y$ that minimizes the sum of the function $f\left(y\right)$ and the distance cost.

We use the proximal operator to compute each generalized directional derivative $f'(x;\mbf{v})$. Given a function $f$ representing a program operation, we constrain the proximal operator from Eq.~\ref{eq:prox_op} to a direction $\mathbf{v}$: 
\begin{align}
    \prox_f(x; \mathbf{v}) &= \arg\min_y \left(f\left(y\right) + \tfrac{1}{2}|| x - y||_2^2\right) \label{eq:dir_prox}\\
    &\trm{where } y=x+t\mathbf{v}: t \in \mathbb{N}, y\in X^n \nonumber
\end{align}
We then define the proximal directional derivative based on the difference with $\prox_f(x; \mathbf{v})$ constrained in the direction $\mbf{v}$ and scaled by the direction magnitude $||\mbf{v}||_2$:
\begin{align}
     \prox'_{f}(x; \mathbf{v}) &= \frac{f(\prox_f(x; \mathbf{v})) - f\left(x\right)}{||\prox_f(x; \mbf{v}) - x||_2} * ||\mbf{v}||_2 
     \label{eq:prox_dir_deriv1}
\end{align}
This takes the same form as the generalized directional derivative (Eq.~\ref{eq:generalized_dir_deriv}), but evaluated with the proximal operator. \abhi{you might consider just calling this a generalized directional derivative like above to avoid new terms: proximal directional derivative, but just a thought} A proximal gradient is defined for a set of direction vectors $V$ like the generalized gradient (Eq. \ref{eq:generalized_gradient}) using proximal directional derivatives:
\begin{align}
     \prox'_f (x; \mathbf{V}) = \left[\prox'_f(x; \mathbf{v}_1),  \ldots,  \prox'_f(x;\mathbf{v}_p)\right]
        \label{eq:prox_grad}
\end{align}
Using proximal gradients allows us to evaluate gradients on discrete operations in programs as if they were continuous nonsmooth functions and apply the associated chain rule for generalized gradients in Eq. \ref{eq:directional_chain_rule_gradient}. \rv{For the rest of this paper, we refer to `proximal gradients' simply as `gradients' unless otherwise specified.}
\abhi{this is great, below you also say generalized derivative and proximal derivative but maybe just clarify this means generalized directional derivative }
\subsection{Program Gradient Evaluation}
\label{sec:method:program_grad_eval}

To compute gradients over programs with PGA, we model a program as a discrete function $P:X^n \rightarrow X^n$, and model the program state $x \in X^n$ as a vector (e.g. $x$ could model a byte array of size $n$ representing the program memory and registers). $P$ is composed of $N$ functions $P_i:X^n \rightarrow X^n, i\in\{1..N\}$ representing individual operations on the program state: 
\begin{align*}
P(x) &= P_N \circ P_{N-1} \circ \cdots \circ P_2 \circ P_1 (x)
\end{align*}
Each program operation $P_i$ is modeled as a combination of $n$ non-smooth scalar valued functions $f_{ij}:X^n \rightarrow X, j\in\{1..n\}$ that define how $P_i$ modifies each variable in the program state. 
\begin{align*}
P_i(x) &= \big[f_{i,1}(x), f_{i,2}(x), ..., f_{i,n}(x)\big] 
\end{align*}
We evaluate each $f'_{ij}$ in $P'_i$ using the proximal directional derivative (Eq. \ref{eq:prox_dir_deriv1}):
\begin{align*}
    P'_i(x; \mbf{v}) = \big[\prox'_{f_{i,1}}(x; \mbf{v}),..., \prox'_{f_{i,n}}(x; \mbf{v}) \big]
\end{align*}

To compose derivatives for a given operation $P_i$ from the previous operation $P_{i-1}$, we individually compose the derivatives of each $f_{ij}$ in $P_i$ from the previous operation $P_{i-1}$:
\begin{align}
    (P_i &\circ P_{i-1})'(x; \mathbf{v}) = \label{eq:P_chain_rule}\\ 
    &\big[(f_{i,1} \circ P_{i-1})'(x; \mathbf{v}),..., (f_{i,n} \circ P_{i-1})'(x; \mathbf{v}) \big] \nonumber
\end{align}
where each $(f_{ij} \circ P_{i-1})'(x; \mathbf{v})$ is defined based on the directional derivative chain rule in Eq.~\ref{eq:directional_chain_rule_gradient}:
\begin{align*}
    (f_{ij} \circ P_{i-1})'(x; \mathbf{v}) = f'_{ij}\left( P_{i-1}(x); P'_{i-1}(x; \mathbf{v}) \right)
\end{align*}
Using the chain rule from Eq.~\ref{eq:P_chain_rule}, we can compute a directional derivative for each final state of the program $P$
by chaining derivatives of the individual operations.
\begin{align*}
P'(x;\mbf{v}) &= (P_N \circ P_{N-1} \circ \cdots \circ P_2 \circ P_1)' (x; \mbf{v})
\end{align*}
We then compute the proximal gradient using Eq.~\ref{eq:prox_grad} for each program state by combining derivatives for a set of direction vectors represented by a matrix $\mbf{V}$:
\begin{align*}
P'(x;\mbf{V}) &= \big[ P'(x;\mbf{v}_1), P'(x;\mbf{v}_2),...,P'(x;\mbf{v}_p)  \big]
\end{align*}

This is the same approach used in Automatic Differentiation, but extended to discrete functions and generalized gradients.
This chained gradient approximation is designed to be error-free for all locally Lipschitz convex functions as well as some locally Lipschitz non-convex functions that meet the requirements for the non-smooth chain rule (e.g., monotonicity).  
\abhi{I would not write something like this, too strong of a theoretical statement. I see you also removed our original proof due to the modified proximal operator, good call}

\subsection{Proximal Derivative Evaluation}
\label{sec:method:prox_dir_derivs}

\abhi{in Figure 4, maybe use the directional derivative notation $f'(x; d)$, not $f'(x)$. Also, using two arrows is kinda confusing cause its unclear which is the gradient in Figure 5 and 4. Lastly, you may want to align the axes, it's -4 to 4 and 0 to 8 in Figure 5 and 4}

\begin{figure}
  \centering
  \captionsetup[subfigure]{indention=17pt}
  \begin{subfigure}[t]{0.45\columnwidth}
    \includegraphics[width=\linewidth]{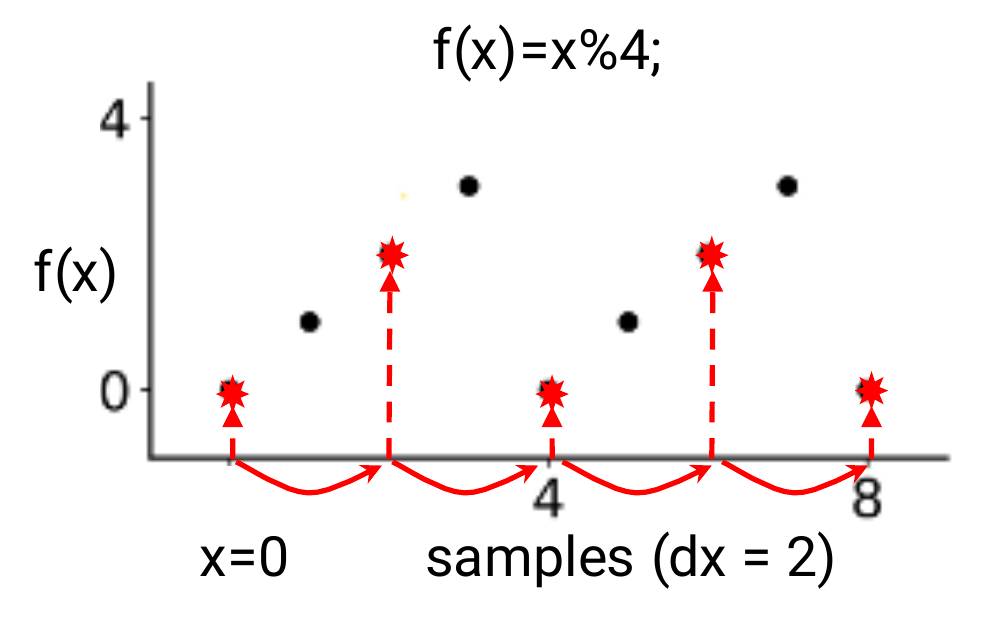}
    \caption{\label{fig:sampling1}{\rm Sample with derivative \tc{dx/dinput=2}.}}
  \end{subfigure}
  \hspace{0.1cm}
  \begin{subfigure}[t]{0.49\columnwidth}
    \includegraphics[width=\linewidth]{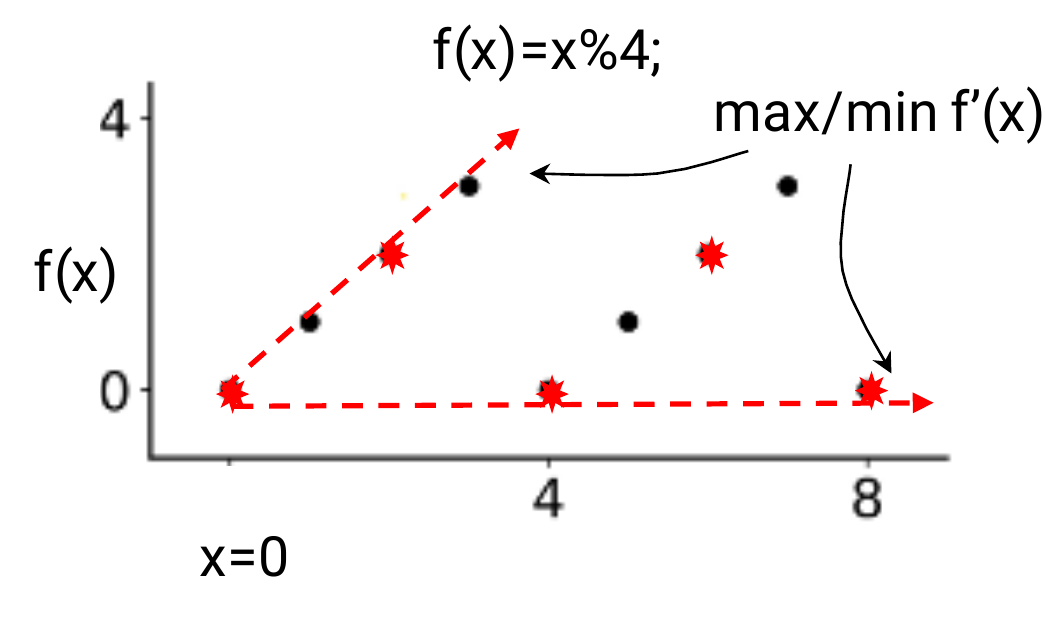}
    \caption{\label{fig:sampling2}{\rm Compute min/max $f'(x)$ from samples.}}
  \end{subfigure}
  \vspace{-4pt}
  \caption{\label{fig:sampling}
  \rvv{Derivative sampling procedure on an \tc{x\%4} operation where the \tc{x} derivative wrt. input \tc{dx/dinput=2}. Samples are first collected at intervals of 2 and then used to compute the max/min directional derivative.}
  }
\end{figure}

\begin{figure}
  \centering
    \captionsetup[subfigure]{indention=17pt}
  \begin{subfigure}[b]{0.47\columnwidth}
    \includegraphics[width=\linewidth]{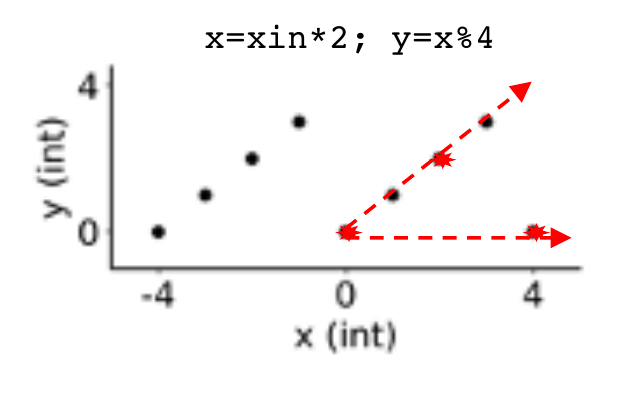}
               \setlength{\abovecaptionskip}{-10pt}
    \caption{\label{fig:working_ex1}{\rm Composition of \tc{mul} 2 with \tc{mod} 4. When \tc{dx/dxin=2}, \tc{dy/dx=1}.}}
  \end{subfigure}
  \hspace{0.1cm}
  \begin{subfigure}[b]{0.47\columnwidth}
    \includegraphics[width=\linewidth]{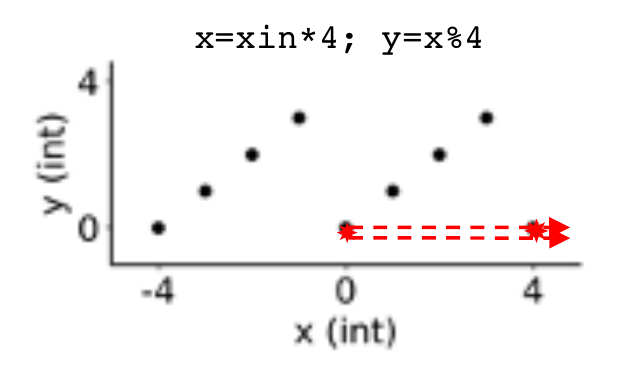}
           \setlength{\abovecaptionskip}{-10pt}
    \caption{\label{fig:working_ex2}{\rm Composition of \tc{mul}  4 with \tc{mod} 4. When \tc{dx/dxin=4}, \tc{dy/dx=0}.}}
  \end{subfigure}
  \vspace{-4pt}
       \setlength{\belowcaptionskip}{-10pt}
  \caption{\label{fig:working_ex}\rvv{
  Proximal Derivative evaluation on composition of a \tc{mul} and \tc{mod} operation at \texttt{xin=0}, with samples  in {\color{red} red}. 
  The step size for the proximal derivative on \tc{x\%4} is determined by the derivative \tc{dx/dxin}. 
  In subfigure (b), when \tc{xin} is first multiplied by 4, \tc{dx/dxin=4} and the sample step size for \tc{x\%4} is 4. This causes the proximal derivative to evaluate to 0, which correctly indicates there is no dataflow over \tc{x\%4} after \tc{mul} by 4.
  }
  }
\end{figure}

When applying proximal directional derivatives in practice, we make two modifications to the proximal directional derivative defined in Eq.~\ref{eq:prox_dir_deriv1} to model program behavior more closely. 

First, 
we only consider the inputs to the operation itself in a function $f^*:X^k \rightarrow X, k \in \{1..n\}$ and associated $v^*$, where $k$ is the number of inputs to the operation. 
To simplify notation we drop the $*$, and for the rest of the paper assume $f$ and $\mbf{v}$ to refer to their $k$ dimensional variants on the current operation.

Second, we modify the proximal operator to select a nearby point that maximizes {\it absolute change} in $f$, which we denote $|\delta f |$:
\begin{align}
\prox_{|\delta f|}(x; \mbf{v})& = \arg \min_y \left(-|f(x)-f(y)| + \tfrac{1}{2}||x-y||_2^2 \right) \label{eq:prox_dir_abs}\\
&\trm{where } y=x+t*\mathbf{v}: t \in \mathbb{N}, y\in X^n \nonumber
\end{align}
This modified proximal operator selects the largest generalized derivative of $f$ based on either the maximum or minimum of $f$ in the direction $\mbf{v}$ (these correspond to the {\it supremum} or {\it infinum} of a generalized derivative). \abhi{consider a footnote instead; also infinum correspond to a set so not clear wht infinum of a generalized derivtive is, maybe set of generalized derivatives? may consider ignoring max/min argument?}
Accounting for both is necessary in dataflow analysis to avoid missing possible dataflows. 
\abhi{you may consider putting your nice image in an overview section with a code snippet!}



\begin{algorithm}[t]
\caption{\textbf{Proximal Derivative computation on a non-smooth operation.}}
\label{alg:prox_deriv}
\lstset{basicstyle=\ttfamily\footnotesize, breaklines=true}
\begin{tabular}{|lrp{1.9in}|}\hline
\textbf{Input}:
    & $\op $&$\leftarrow$ program operation \\
    & $x1, x2 $& $\leftarrow$ operation inputs \\
    & $dx1, dx2 $ & $\leftarrow$ $x1, x2$ components of $\mbf{v}$ \\
    & $N $ & $\leftarrow$ maximum samples \\
\hline
\end{tabular}

\begin{algorithmic}[1]
\IF{$dx1 = 0$ \AND $dx2 = 0$}
    \RETURN $dy \leftarrow 0$
\ENDIF
\STATE $y \leftarrow \op(x1, x2)$
\STATE initialize size $N$ arrays $S$ and $S_{cost}$
\FOR{$i = 1$ to $N$}
    \STATE $x1_{i} \leftarrow x1 + dx1*i$
    \STATE $x2_{i} \leftarrow x2 + dx2*i$
    \STATE $y_i \leftarrow \op(x1_i, x2_i)$
    \STATE $distance_i^2 \leftarrow (x1 - x1_i)^2 + (x2 - x2_i)^2$
    \STATE add $-|y - y_i| + \tfrac{1}{2}distance_i^2$ to $S_{cost}$ array
    \STATE add $y_i$ to $S$ array
\ENDFOR
\STATE $i_{prox} \leftarrow$ index of min sample in $S_{cost}$
\STATE $y_{prox} \leftarrow $ recover sample $i_{prox}$ from $S$
\RETURN $dy \leftarrow (y_{prox} - y)/i_{prox}$
\end{algorithmic}
\end{algorithm}

\parheader{Proximal Derivative Algorithm.} 
Algorithm \ref{alg:prox_deriv} defines how we compute the proximal derivative for an operation $\op$ that has two input variables $x1$ and $x2$, and returns an output $y$.  We denote the derivatives of the inputs $x1$ and $x2$ and output $y$ to be $dx1$, $dx2$, and $dy$, where $dx1$ and $dx2$ are components of $\mbf{v}$, and $dy$ is computed using the proximal derivative with a maximum sample budget $N$. The same algorithm can be applied to operations with any number of inputs from $1$ to $n$ by adjusting the number input variables.
\abhi{maybe change $S_{cost}$ to Cost; also the Algorithm doesn't match equation 8 anymore? in partcicular, there is no division anymore by a distance and no scale by direction magnitude?}
Figure \ref{fig:sampling} shows an example of the proximal derivative procedure being applied to a \tc{x\%4} operation.

We observed that when the proximal gradient is nonzero, it almost always uses a point within a few samples of the current point due to rapid increase of the proximal cost term $distance^2$ in the proximal operator. Therefore, we set $N$ to a small constant (5 in our evaluation), and evaluate the proximal derivative in that range.

Figure \ref{fig:working_ex} gives an example of evaluating Algorithm \ref{alg:prox_deriv} on a non-smooth operation \texttt{y = x\%4}. When the input is multiplied by 2 as in Figure \ref{fig:working_ex1}, the algorithm samples at intervals of 2 and evaluates a derivative of 1 based on the maximum absolute difference ($|\delta f|$) measure. However, when the input is multiplied by 4 as in Figure \ref{fig:working_ex2}, the algorithm samples at intervals of 4 and evaluates a derivative of 0 because the samples are all 0. This 0 derivative indicates that the composition of functions \texttt{x=xin*4; y=x\%4} will always have the same output and therefore has no dataflow.

\subsection{Derivative Propagation Rules} 
\label{sec:methodology:prop_rules}
We define a general framework for propagating derivatives over 5 abstract classes of operations that need to be handled in program analysis: floating point operations, integer valued operations, loading and storing variables, branching, and function calls to external libraries. 

\begin{enumerate}
    \item {\bf Floating point operations}: We treat floating point operations as continuous \abhi{differentiable i believe, not continuous} functions and apply the standard chain rule (Eq.~\ref{eq:P_chain_rule}) with their analytic derivatives. If there are any potentially non-smooth floating point operations, such as floating point modulo, or typecasting between floating point types, we use proximal derivatives.
    
    \item {\bf Integer operations}: We consider any boolean or typecasting involving integers to be integer operations, as well as any arithmetic, bit shifting, or modulo on integer or pointer types. In general we use proximal derivatives on all integer operations, although in some cases such as arithmetic addition and multiplication we use analytic derivatives as an optional optimization.
    
    \item {\bf Load and Store}: When variables are stored or loaded from memory, their associated derivatives are also stored or loaded (our implementation uses shadow memory to track derivatives in memory, although any associative tracking mechanism could be used). If the memory address passed to a load instruction has a nonzero derivative, we set the derivative of the loaded variable to 1.0 if it does not already have a nonzero derivative. 
\rv{    This is a simplifying approximation that may lead errors in evaluating the proximal gradient. However, we note that proximal derivatives on load operations can potentially be evaluated by sampling adjacent memory locations. We leave this to future work.}

    \item {\bf Branches}: When dynamically computing derivatives, we can only reason about the derivative on the current execution path. If computing a derivative would require sampling an alternate execution path, we instead set that derivative to 0. Therefore, when a branch is encountered, we set any derivatives to 0 that are based on samples that would change the branch condition. 
\rv{    This approach may miss some parts of the gradient but ensures we do not propagate incorrect derivatives. We note that sampling across multiple execution paths when handling branches could yield more accurate proximal derivatives and reason about control flow data flows (i.e. implicit data flows), we leave this to future work.}
    
    
    
    
    \item {\bf External Library Functions}: Provided they do not have side effects, derivatives on external library function calls can be computed using proximal derivatives, while functions with side effects must be handled on a case by case basis. When an external function overwrites a buffer, we also clear the stored derivatives associated with that buffer.
\end{enumerate}

\subsection{Program Gradient as Dataflow} 
\label{sec:method:program_grad_dataflow}
To use gradients as a measure of dataflow, we compute gradient between a set of user defined sources and sinks. We set the initial vectors in $\mbf{V}$ so that each vector is all $0$s except for an initial derivative on each source of $+1$ or $-1$ \abhi{maybe explicitly state +1, -1 is a direction}. We then execute the program and propagate the derivatives over each operation with the chain rule and derivatives defined in Algorithm~\ref{alg:prox_deriv}. While the program is executing we record derivatives at each sink, and accumulate the gradient on each sink from all the sources. Cumulatively, the gradients on all sinks form the Jacobian $\mathcal{J}$ between sources and sinks. \abhi{I think derivative and gradient are mixed here}

Algorithm \ref{alg:gradient_dataflow} formally describes the process for computing the gradients from a set of sources to each designated sink in program. The returned Jacobian $\mathcal{J}$ contains the gradients of each sink based on the largest derivative propagated to it from each source \rv{(sinks may record multiple derivatives from a single source if, for example, the sink is in a loop)}.

\begin{algorithm}
\caption{\textbf{Program Gradient Evaluation.}}
\label{alg:gradient_dataflow}
\lstset{basicstyle=\ttfamily\footnotesize, breaklines=true}
\begin{tabular}{|lrp{1.9in}|}\hline
\textbf{Input}:
    & $P $&$\leftarrow$ program under analysis \\
    & $x $&$\leftarrow$ program input \\
    & $Sources $& $\leftarrow$ $n$ dataflow sources \\
    & $ Sinks $& $\leftarrow$ $m$ dataflow sinks \\
\hline
\end{tabular}

\begin{algorithmic}[1]
\STATE initialize $\mbf{V}$ to empty set $\{\}$
\STATE initialize $\mathcal{J}$ $n \times m$ matrix to $0$s
\FOR{ $src$ in $Sources$ }
    \STATE $\mbf{v}^+_{src} \leftarrow [dsrc=1, \trm{otherwise }0]$
    \STATE $\mbf{v}^-_{src} \leftarrow [dsrc=-1, \trm{otherwise }0]$
    \STATE add $\mbf{v}^+_{src}$ and $\mbf{v}^-_{src}$ to $\mbf{V}$
\ENDFOR
\STATE Execute $P$ on input $x$, tracking $P'(x; \mbf{V})$
\FOR{ $sink$ in $Sinks$}
    \FOR{each recorded $\tfrac{dsink}{dsrc}$}
        \IF { $\left|\tfrac{dsink}{dsrc}\right| > \big| \mathcal{J}[src, sink] \big|$ }
            \STATE $\mathcal{J}[src, sink] \leftarrow \tfrac{dsink}{dsrc}$
        \ENDIF
    \ENDFOR
\ENDFOR
\RETURN $\mathcal{J}$
\end{algorithmic}
\end{algorithm}

\section{Implementation}
\label{sec:implementation}

We implement PGA as a new sanitizer in the LLVM framework~\cite{llvm2004} called Gradient Sanitizer (\texttt{grsan}). We use LLVM because it allows us to instrument a program during compilation after it has been converted to LLVM's intermediate representation. This means that \texttt{grsan} can be used to instrument any program written in a language supported by LLVM, and incurs lower runtime overhead than binary instrumentation frameworks such as PIN or Valgrind \cite{pin, nethercote2007valgrind}. \rrevise{However, we note that PGA could also be implemented in a binary instrumentation framework to facilitate analysis in cases where source code is not available.}


\noindent \textbf{Overall Architecture.} We base \texttt{grsan} on LLVM's taint tracking implementation, DataFlowSanitizer (\texttt{dfsan}), which uses shadow memory to track taint labels. 
For each byte of application memory, there are two corresponding bytes of shadow memory that store the taint label for that byte.  

\begin{figure}
    \includegraphics[width=\columnwidth]{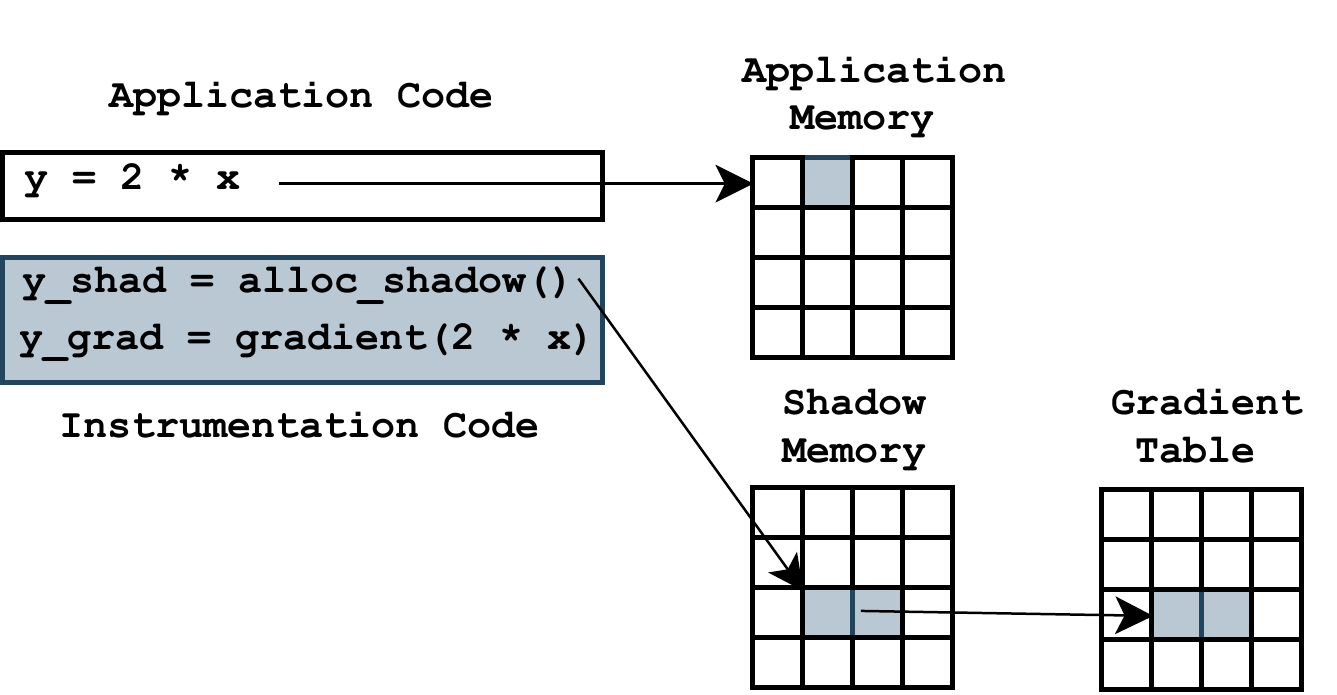}
  \vspace{-15pt}
  \caption{\label{fig:grsan_diagram} \texttt{Grsan} architecture illustrating how proximal gradients are propagated.}
  \vspace{-10pt}
\end{figure}

We modify \texttt{dfsan} in the following two ways: First, we add additional metadata associated with each label that stores the gradient information, which is stored in a separate table as shown in Figure~\ref{fig:grsan_diagram}. Each label in the shadow memory is associated with a distinct derivative value in the gradient table. 
The 0 label is reserved for 0 derivative, and any shadow memory lookup on a constant or unlabeled variable returns label 0. 

Second, we change the dataflow propagation rules to compute gradients over each operation. Figure \ref{fig:grsan_diagram} shows an example of how the \texttt{grsan} instrumentation works. Given an operation \tc{y=2*x}, the instrumentation first looks up the derivative for each input, \tc{2} and \tc{x}, from shadow memory.  If any input has a nonzero derivative, it computes the derivative for the output \tc{y} and generates a new shadow memory label by incrementing the current max label by 1. It then allocates space in the shadow memory and gradient table and stores the new label and associated derivative of \tc{y}. 

As an additional optimization, when storing an operation's output derivative we first compare it to the input derivatives. If the output derivative is equal to either, we apply the label of the equivalent input derivative to the output instead of generating a new label and gradient table entry. Since many operations do not change the value of the derivative (e.g. \tc{x = x+1;}), this significantly reduces the number of distinct labels that need to be tracked.

\rv{In the current implementation, \texttt{grsan} tracks derivatives from a single source at a time, propagating the two derivatives \abhi{maybe say refer to the figure why it is 2 derivatives} from the source in parallel. When computing a gradient over multiple sources (e.g. bytes in an input file), we execute the program once for each source. We intend to extend \texttt{grsan} to support multiple sources in parallel in future work.} \abhi{you might also write we use a python script to parallelize it?}

\noindent \textbf{Gradient Propagation Instrumentation.} 
For differentiable operations such as a floating point multiplication (\tc{fmul}), \rv{\texttt{grsan} uses the analytical derivative of the operation}. \abhi{technically according to section 3, this is only done for float types?}For nondifferentiable operations such as bitwise \tc{And}, \texttt{grsan} uses an optimized version of proximal derivatives from Algorithm \ref{alg:prox_deriv} that returns the first nonzero derivative it encounters when sampling. \abhi{maybe connect to earlier discussion of fixed budget sampling} 
We found this approximation picked the same values that the proximal operator would select and is computationally lighter (i.e. does not require computing exponents). 




We leave most external function calls uninstrumented, but some operations in \texttt{glibc} are given special instrumentation. We set the gradients for any buffer overwritten by \texttt{fread} or \texttt{memset} to 0, and the gradients of buffers copied by \texttt{memcpy} or \texttt{strcpy} are also copied. Type casting instructions are handled by simply copying labels from the original value to the result.

\section{Evaluation}
\label{sec:evaluation}

We evaluate PGA by comparing its performance directly to DTA, and in direct applications for bug finding and security analysis. Specifically, we run experiments to answer the following questions:

\begin{enumerate}[topsep=0pt,itemsep=-1ex,partopsep=1ex,parsep=2ex]
  \item \tbf{Dataflow Accuracy:} Is PGA more accurate than DTA in tracking dataflows?
  \item \tbf{Overhead:} How does the overhead introduced by PGA compare to DTA?
  \item \tbf{Guided Fuzzing:} Does using PGA to guide fuzzing lead to better edge coverage?
  \item \tbf{CVE Analysis:} Can PGA detect and analyze recent CVEs that taint is typically used to detect?
  \item \tbf{Bug Discovery:} Is PGA an effective tool for finding bugs?
  \item \tbf{Information Leaks:} Can PGA detect and analyze memory and timing-based information leaks?
\end{enumerate}

\subsection{Experimental Setup}

\noindent \textbf{Test Programs.} We perform tests on a set of 5 widely used file parsing libraries and 7 total programs. We use file parsers because these programs often must process files from untrusted sources, making them a common target for attacks. Table~\ref{tab:programs} shows the test programs and SLOC associated with each executable tested. In total the programs have 391,883 SLOC. 


\parheader{Fuzzers Evaluated.} For our fuzzing experiments, we use the latest version of \texttt{NEUZZ}~\footnote{www.github.com/Dongdongshe/neuzz} and \texttt{VUzzer}~\footnote{www.github.com/vusec/vuzzer64}.

\noindent \textbf{Test Environment.} All of our evaluations are  performed on an Ubuntu 16.04 server with an Intel Xeon E5-2623 v4 2.60GHz CPU and 192G of memory unless otherwise specified.

\begin{table}
\centering
{\small
 \begin{tabular}{ll l l} 
 \toprule
  Library & Test Command & SLOC & File Format \\ 
 \midrule
 zlib-1.2.11    & \verb|minigzip -d| &  3228 & GZ/ZIP \\ 
 libjpeg-9c     & \verb|djpeg| & 8,857    & JPEG \\
 mupdf-1.14.0   & \verb|mutool show| & 123,562    & PDF \\  
 libxml2-2.9.7  & \verb|xmllint| & 73,920    & XML \\
 binutils-2.30 & \verb|objdump -xD| & 72,955  & ELF \\  
                & \verb|strip | &  56,330   &  \\
                & \verb|size| & 52,991    & ELF \\  

 \bottomrule
 \end{tabular} }
          \setlength{\belowcaptionskip}{-10pt}
  \caption{\label{tab:programs} Test programs used in our evaluation.}
  \vspace{-0pt}
\end{table}


\begin{table*}
\small{
\begin{tabular}{l | ccc | ccc | ccc | ccc | ccc}
\toprule


\multicolumn{1}{c}{ } & \multicolumn{3}{c}{{\bf Neutaint}} & \multicolumn{3}{c}{{\bf libdft}} & \multicolumn{3}{c}{{\bf dfsan}} & \multicolumn{3}{c}{{\bf grsan (binary)}} & \multicolumn{3}{c}{{\bf grsan (floats)}} \\
&Prec.&Rec.&F1&Prec.&Rec.&F1&Prec.&Rec.&F1&Prec.&Rec.&F1&Prec.&Rec.&F1\\
\midrule
minigzip    &0.02&0.55&0.04    &0.42&0.29&0.17     &0.29&0.60&0.39 &0.41&0.15&0.22&
0.63&0.51&\tbf{0.57}\\
djpeg       &0.02&0.33&0.04    &-&-&-&  0.22&1.00&0.37 & 0.62&0.63&0.62&
0.60&0.83&\tbf{0.69}\\
mutool      &0.002&0.19&0.004&0.70&0.32&0.22     &0.63&0.61&0.62 &0.87&0.50&0.63&
0.86&0.51&\tbf{0.64}\\
xmllint     &0.07&0.69&0.12     &-&-&-              &0.62&0.99&0.76 &0.91&0.87&0.89&
0.94&0.91&\tbf{0.92}\\
objdump     &0.03&0.20&0.05     &0.47&0.67&0.28     &0.37&0.93&0.52 &0.51&0.66&0.58&
0.66&0.77&\tbf{0.71}\\
strip       &0.02&0.39&0.03     &0.26&0.59&0.18     &0.20&0.96&0.33 &0.42&0.72&0.53&
0.50&0.86&\tbf{0.63}\\
size        &0.06&0.39&0.11     &0.20&0.59&0.30     &0.37&0.95&0.53 &0.54&0.76&0.63&
0.62&0.91&\tbf{0.74}\\
\bottomrule
\end{tabular}
}
 \caption{\label{tab:precision_comp}Summary of accuracy comparison results for DTA and PGA systems. \texttt{Neutaint}, \texttt{libdft}, and \texttt{dfsan} are state-of-the-art DTA systems, while binary \texttt{grsan} is an ablation of PGA that only uses binary (1 or 0) gradients to test the impact of precise gradients on accuracy. Best F1 scores for each program are highlighted. Experiments with \texttt{libdft} on \texttt{djpeg} and \texttt{xmllint} timed out after 24hrs. PGA (with floating point gradients) outperforms DTA on all programs, and full precision (floats) \texttt{grsan} outperforms binary \texttt{grsan} on all programs.}
 \vspace{-10pt}
\end{table*}

\subsection{Performance}

We first evaluate the performance of PGA as a tool for dynamic dataflow analysis. In our experiments, we compare PGA to DataFlowSanitizer (\texttt{dfsan}), LLVM's state-of-the-art DTA implementation. Since our implementation of PGA is based on the \texttt{dfsan} architecture, our setup ensures that any differences in performance between PGA and DTA are to due the respective performance of gradient and taint and not due to differences in the underlying architectures. 

We compare performance in three areas: first, we estimate the accuracy of the dataflows predicted by PGA and DTA. Second, we evaluate the overhead introduced by the PGA  instrumentation. Third, we compare the edge coverage achieved by a dataflow-guided fuzzer using either PGA or DTA to guide its mutation strategy. 

\parheader{Evaluation Inputs.} \rv{We use the same set of initial input files for all of the performance evaluations. The gzip, pdf, and ELF files are sourced from the AFL sample seeds included in the distribution\footnote{https://github.com/google/AFL}. The jpeg input was generated from running a small jpeg image through a jpeg reduction service\footnote{https://tinyjpg.com/}. The libxml input was selected from the libxml\footnote{https://gitlab.gnome.org/GNOME/libxml2/} test inputs smaller than 700 bytes with the greatest AFL branch coverage.  }

\subsubsection{Dataflow Accuracy}
\label{sec:accuracy}


We evaluate the accuracy of PGA in comparison to DTA against an estimate of ground truth dataflows. This comparison setting favors DTA since it does not take the fine grained dataflow information from PGA into account (i.e., only considers binary 0/1 influence), but still illustrates the benefits of PGA's increased precision. \revise{In addition to comparing against \texttt{dfsan}, we also compare against \texttt{libdft}, another widely used DTA framework that uses Intel PIN to instrument the binary directly, \texttt{Neutaint}, which uses the gradients of a neural network to model dataflows, and an ablation of PGA with binary gradients, \texttt{grsan (binary)}. Notably, \texttt{libdft} tracks taint at byte level granularity and incorporates special case rules to handle operations that cancel out dataflows, such as \tc{y = x - x}.}

\noindent \textbf{Ground truth estimation.} To estimate ground truth dataflows, we measure if changes in taint sources cause changes in sink values during execution. When recording executions, we only consider executions that follow the same path to remove implicit flows, since neither DTA nor PGA can detect these.
We mark each byte read from the input file as a source and each branch condition as a sink, because branches ultimately determine the behavior of a program, and because many security vulnerabilities can only be exploited when certain branches are taken. 
For each input byte, we set the byte to 0, 255, and toggling each bit for a total of 10 samples. We found that this sampling strategy usually triggered a change in the sink variable when there was a valid dataflow. \abhi{maybe explicitly say this is not perfect ?  i think people complained before}



\noindent \textbf{Accuracy evaluation.} We perform the accuracy evaluation on the programs shown in Table~\ref{tab:programs} using a set of small seed files (<1Kb) to make sampling each byte feasible. 
Since valid dataflows often only involve a few input bytes, 
we use F1 accuracy, which is a standard metric for evaluating predictions on imbalanced classes in classification problems. F1 accuracy is computed as 
$\textrm{F1}=2*\tfrac{\textrm{precision}*\textrm{recall}}{\textrm{precision}+\textrm{recall}}$. Precision indicates the proportion of bytes with predicted dataflows that are correct (i.e. not false positives), while recall indicates the proportion of valid dataflows that were correctly predicted (i.e. not false negatives). Results are shown in Table~\ref{tab:precision_comp}.

Generally, PGA achieves a significant improvement in precision, achieving up a 37\% increase in precision and 33\% increase in F1 accuracy (20\% on average) compared to the best performing DTA system, \texttt{dfsan}. Overall PGA gets higher F1 scores for all programs. In spite of incorporating special case dataflow cancellation rules for its bitwise and numerical operations, \texttt{libdft} achieves lower accuracy than \texttt{dfsan} in the evaluation. We hypothesize this is due to the difficulty in writing handcrafted rules for all possible X86 instructions, which leads to errors in propagation rules as noted in ~\cite{taintinduce}. The binary gradient PGA ablation, \texttt{grsan (binary)}, also has much lower accuracy than full precision PGA, indicating gradients are essential to computing accurate dataflows with PGA. We discuss the binary gradient ablation in more detail in Appendix \ref{app:binary_grad_ablation}.

\vspace{1pt}
\noindent \begin{longfbox}
{\tbf{Result 1:} PGA achieves the highest F1 accuracy on all 7 tested programs compared to 3 state-of-the-art DTA systems, and is up to 33\% more accurate than the next most accurate DTA system, \texttt{dfsan}.}
\end{longfbox}\\

\begin{figure*}[ht!] 
  \centering
  \includegraphics[width=\linewidth]{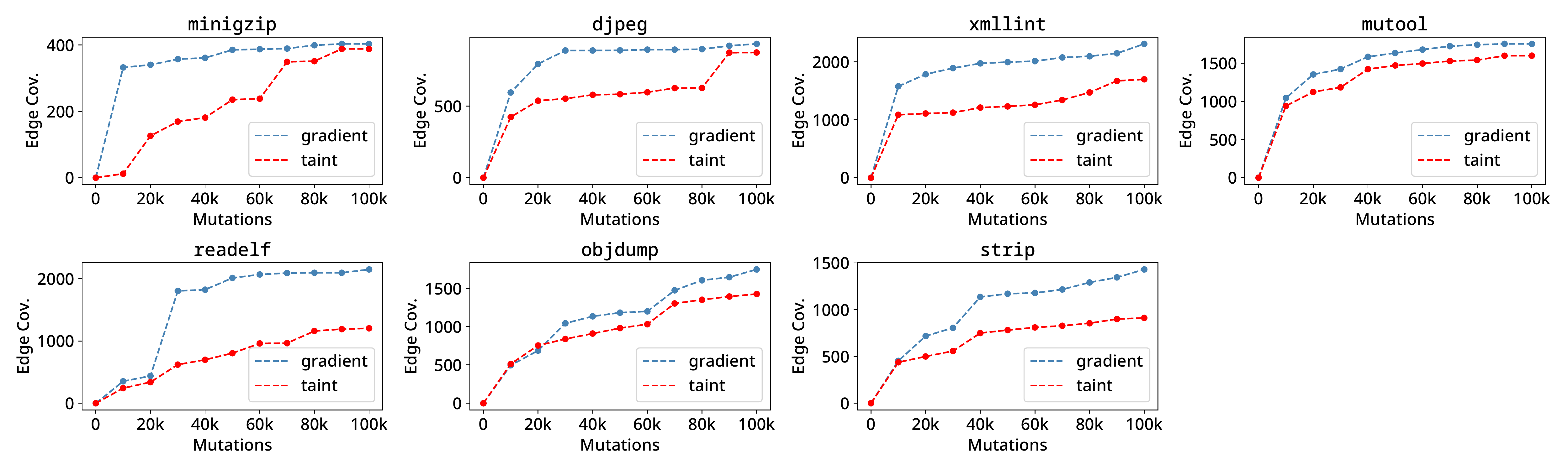}
  \vspace{-15pt}
         \setlength{\belowcaptionskip}{0pt}
   \caption{ \label{fig:fuzz_comp}Comparison of guided fuzzer edge coverage achieved by PGA and DTA over 100k mutations from a single seed. Overall gradient-guided fuzzing achieves up to 56\% higher coverage and improves the rate of new edge discovery by 10\% on average. }
  \vspace{-10pt}
\end{figure*}

\parheader{Additional Accuracy Experiments.} In addition to the accuracy experiment in Table \ref{tab:precision_comp}, we run experiments to address the following: (1) How do varying compiler optimization levels effect the accuracy of PGA vs. DTA? (2) How does PGA perform against \texttt{Neutaint} in Hotbyte prediction? (3) On which specific operations does PGA vary from DTA due to 0 gradients? (4) How does PGA compare with Quantitative Information Flow (QIF) techniques? We summarize the results here and describe these experiments in detail in Appendix \ref{app:additonal_acc_exps}.
\begin{enumerate}[topsep=0pt,itemsep=-1ex,partopsep=1ex,parsep=2ex]
    \item {\bf Compiler Optimization.} PGA's accuracy improvement over DTA is robust to varying compiler optimization levels. On average, PGA is at least 18\% more accurate than DTA on compiler optimization levels \texttt{-O0} through \texttt{-O2}.
    \item {\bf Hotbyte Prediction.} When we reproduce the Hotbyte experiment described in \texttt{Neutaint} ~\cite{she2019neutaint}, (i.e. identifying input bytes with the most dataflows to branches) PGA achieves 43.8\% accuracy while Neutaint achieves 64.3\% accuracy on average. \rv{Neutaint achieves higher average accuracy because it trains on a large corpus of recorded execution traces, while PGA and the DTA reason about a single input and execution trace at a time. We see \texttt{Neutaint} as a \rvv{complementary} method that performs well in identifying hotbytes, while PGA has better fine grained dataflow accuracy, and both methods could be used together in program analysis.} \abhi{seems like rreviewer wanted more detail here, likely with respect to Accuracy Evalatuion in 5.2.1?}
    \item {\bf Zero Gradient Analysis.} PGA avoids overtainting errors when it computes zero gradients on operations where DTA would propagate taint. We find the zero gradients occur most frequently on And, Remainder, Sub, Mul, and Shift operations, and that zero gradients are most often caused by masking, shifting, or composition effects.
    \item {\bf QIF Comparison.} We compare PGA with a QIF tool \texttt{Flowcheck} that quantifies information flow in the form of bit leakage~\cite{mccamant2008quantitative}. PGA outperforms \texttt{Flowcheck} by 22\% on average in terms of F1 accuracy. 
\end{enumerate}

\subsubsection{Overhead}

\rrevise{We observe two conflicting phenomena when measuring overhead: PGA can either increase overhead due to the additional floating point storage and computation required by gradients, or decrease runtime and memory overhead when its increased precision reduces unnecessary dataflow tracking operations that use additional computation and shadow memory.}

\rv{We evaluate the overhead introduced by our implementation of PGA in runtime and memory relative to \texttt{dfsan} on a single source dataflow. Note that if we consider overhead for multiple sources, the runtime will be lower and the memory overhead will be higher for a multi-source implementation.}
In the worst case PGA has 21.7\% greater overhead in runtime and 21.5\% in memory relative to DTA, but on average only adds 3.21\% relative overhead in runtime and 1.48\% in memory.  Table~\ref{tab:performance} and Table~\ref{tab:memory} in Appendix \ref{app:overhead_evaluation} show the detailed results. We also provide overhead measurements for \texttt{libdft}, although it adds significantly more overhead due to the binary instrumentation.


\vspace{1pt}
\noindent \begin{longfbox}
{\tbf{Result 2:} On average PGA increases runtime overhead by 3.21\% runtime and memory overhead by 1.48\% relative to DTA, and increases runtime by 21.7\% and memory usage by 21.5\% relative to DTA in the worst case. }
\end{longfbox}

\subsubsection{Dataflow-Guided Fuzzing}
\label{sec:fuzzing_eval}

Since dynamic dataflow analysis is often used as a tool to guide fuzzing, we evaluate PGA in comparison to DTA as a method for guiding fuzzer mutations. Unlike our evaluation of dataflow accuracy, this experiment emphasizes the dataflow magnitude information provided by the program gradient, since bytes with the largest derivatives are selected for fuzzing. 

\rvv{We first compare PGA and DTA using a simple deterministic strategy for mutating input bytes based on dataflows to branches. This ensures there is no bias from randomized mutation strategies or other heuristics employed by state-of-the-art fuzzers in this evaluation.}
First, we execute the program with all inputs set as sources and all branches set as sinks. We then select 128 bytes from the input bytes based on the measured taint and gradient flows to branches. With PGA, the bytes with the greatest gradients are prioritized, this approach utilizes the additional information provided by PGA to improve the mutation strategy. 
The fuzzer performs a deterministic set of mutations on the selected 128 bytes, in which each byte in turn is set to all 256 possible values. 




\noindent \textbf{Edge coverage comparison.} We execute the fuzzer with both PGA and DTA for 100,000 mutations, and record coverage every 10,000 mutations. Figure ~\ref{fig:fuzz_comp} shows the relative edge coverage achieved by each method. On average the gradient guided fuzzing outperforms taint in increasing edge coverage by 10\% per 10,000 mutations. The gradient guided fuzzer achieves higher coverage on all programs, with the greatest improvement in overall edge coverage of 56\% on \texttt{strip}. 
We note that for some programs such as \texttt{xmllint}, there is a significant disparity between the results of the guided fuzzing and precision evaluations. We believe this difference is caused by two factors: the magnitude of the gradient was more important than its accuracy in guiding the fuzzer on these programs, and that even small differences in accuracy can be significant if they allow the fuzzer to precisely target key branches in the program.

\parheader{Enhancing state-of-the-art fuzzers.} We also evaluate if the gradient information from PGA can improve the performance of \texttt{NEUZZ}, a state-of-the-art fuzzer. We evaluate a version of \texttt{NEUZZ} modified to use PGA against unmodified \texttt{NEUZZ} and \texttt{VUzzer}, another dataflow guided fuzzer.  On average, \texttt{PGA+NEUZZ} improves new edge coverage by 12.9\% over baseline \texttt{NEUZZ}. We hypothesize this improvement is because the gradients produced with PGA are more precise than the neural-network based gradients used by by \texttt{NEUZZ}. We discuss this experiment and provide more detailed results in Appendix \ref{app:fuzzing_current_fuzzers}. \abhi{maybe state like in our email, we you cannot do DTA+NEUZZ b/c NEUZZ is gradient-guided}

\vspace{3pt}
\noindent \begin{longfbox}
{\tbf{Result 3:} In guided fuzzing PGA increases the rate of edge coverage growth by 10\% on average compared to DTA, and improves the edge coverage of NEUZZ, a state-of-the-art fuzzer, by 12.9\% on average.}
\end{longfbox}

\subsection{Bug Finding}
\label{sec:eval:bug_finding}

Next, we show the additional information provided by PGA make it a useful tool for discovering and analyzing different types of bugs in real world programs. We test PGA against DTA in three applications: detecting and analyzing known vulnerabilities, guiding discovery of new bugs, and discovering information leaks.

\subsubsection{Analysis of known CVEs}
\label{sec:eval:cves}

We first evaluate PGA as a tool for detecting dangerous dataflows offline in known CVEs.  We instrument the programs to mark user-controlled input as dataflow sources and the instructions involved in the attacks as dataflow sinks. We select 21 CVEs that cover a range of vulnerability types, including stack and heap overflows, integer overflows, memory allocation errors, and null pointer dereferences. We include CVEs from our evaluation programs as well as \texttt{openssl} to demonstrate PGA based analysis on a variety of program types.

Table~\ref{tab:cve_detection} shows a comparison of PGA and DTA in detecting the relevant dataflows in these CVEs. PGA correctly identifies dataflows for 19 out of the 21 evaluated CVEs, including 2 CVEs that cannot be identified with DTA. For these CVEs, DTA overtaints on the malicious inputs and crashes due to label exhaustion, while PGA can precisely identify the dataflows without overtainting. For the 2 CVEs which both PGA and DTA fail to detect, the dataflow source indirectly propagates to the sink through implicit dataflows (i.e. control flow).

We also note the utility of the additional information provided by gradients and how it can help distinguish vulnerabilities in an online manner. 
In the case of CVE-2017-15996, an out of memory allocation error triggered by the dataflow from an input byte, PGA directly measures the effect of input changes on the size of the allocation, and can early terminate when it finds input byte values that will trigger the out-of-memory error. 

\vspace{3pt}
\noindent \begin{longfbox}
{\tbf{Result 4:} PGA identifies relevant dataflows in 19 out of 21 evaluated CVEs, including 2 DTA cannot detect due to label exhaustion. PGA and DTA both cannot identify control-flow-based dataflows for 2 CVEs.}
\end{longfbox}

\begin{table}
\centering
{\small
\begin{tabular}{@{}p{2.38cm} p{4.1cm} p{0.38cm} p{0.4cm} @{}}
\toprule
CVE ID & Vulnerability - Program & PGA & DTA \\
\midrule
CVE-2007-1657  & stack overflow - minigzip &  \checkmark & \checkmark\\

CVE-2017-7210  & off-by-one read - objdump  & \checkmark & \checkmark \\
CVE-2017-8396 & heap overflow - libbfd & \checkmark & \checkmark \\
CVE-2017-15996 & out-of-memory - readelf  & \checkmark  & \checkmark \\ 

CVE-2018-6543 & integer overflow - objdump  & \checkmark & \checkmark\\ 
CVE-2018-6759  & null ptr dereference - nm & \checkmark & \checkmark \\
CVE-2018-7643 & integer overflow - objdump  & \checkmark & \checkmark\\
CVE-2018-10372 & heap overflow - readelf & \checkmark & \checkmark \\
CVE-2018-11813  & infinite loop - cjpeg &  \checkmark & \checkmark\\
CVE-2018-12698 & out-of-memory - libiberty & \checkmark & \checkmark \\
CVE-2018-12699  & heap overflow - libiberty & \checkmark & \checkmark \\
CVE-2020-14152  & out-of-memory - djpeg  & \checkmark & \checkmark \\
CVE-2018-19932 & integer overflow - strip  & \checkmark & \checkmark \\
CVE-2018-19777 & infinite loop - mutool  & \checkmark & \checkmark\\
CVE-2018-20671 & infinite loop - objdump  & \checkmark & \checkmark\\

CVE-2019-14444  & integer overflow - readelf &  \checkmark & \checkmark\\
CVE-2020-1967  & null ptr dereference - openssl  & \checkmark & \checkmark \\

CVE-2018-11212 & divide-by-zero - cjpeg & \checkmark & $\times$\\
CVE-2018-11214 & heap overflow - cjpeg  & \checkmark & $\times$\\

CVE-2020-7041  & invalid certificate - openssl  & $\times$ & $\times$ \\
CVE-2018-12697 & null ptr dereference - libiberty  & $\times$ & $\times$ \\

\bottomrule
\end{tabular}
}
         \setlength{\belowcaptionskip}{-20pt}
\caption{\label{tab:cve_detection}List of 21 CVEs for which the exploitable dataflows were analyzed by PGA and DTA (\texttt{dfsan} and \texttt{libdft})}
\end{table}

\subsubsection{Bug Discovery}

We compare PGA and DTA as bug discovery tools by adding additional instrumentation to record dataflows for instruction and function arguments that can potentially trigger program errors, such as memory allocations, copy instructions, indexing operations, and shift operators. We then execute the programs on a corpus of files generated by running AFL on each program for 24 hours. Next, we generate new inputs by changing input bytes involved in the recorded dataflows similar to Section ~\ref{sec:fuzzing_eval}. For PGA, we select 128 input bytes prioritized based on the function gradient, while for DTA, we randomly select them. We modify the values of the selected bytes based on the gradient for PGA or by setting them to 0 or 255 for DTA.

Table~\ref{tab:bug_summary} summarizes our results. Overall, PGA finds 22 bugs in our evaluated programs \rrevise{through gradient guided modification of the inputs}, including arithmetic errors, out-of-memory allocations, and integer overflows. The DTA guided bug search finds 15 of these 22 bugs. Of the 22 bugs, 20 have been confirmed by the developers, 3 of them resulted in new patches, and the remaining 17 were already patched in the latest sources of the programs.

For the 7 bugs that were found by PGA and not DTA, gradient magnitude and direction allowed the search to prioritize input bytes that could trigger errors that could not be identified with DTA. We give a case study in Figure~\ref{lst:implicit_control}, which illustrates how large gradients are used to find an arithmetic error in \texttt{djpeg}. By altering an input byte with a large gradient to a shift operand, an overflow is triggered that results in an invalid operation. Similarly, identifying inputs with large gradients to memory operations was key to finding memory errors. 

\vspace{3pt}
\noindent \begin{longfbox}
{\tbf{Result 5:} A simple PGA guided search finds 22 bugs in the tested programs. A DTA guided search using the same strategy and inputs finds 15 of these 22 bugs.}
\end{longfbox}

%

\begin{figure}
    \centering
    \begin{lstlisting}
GRSAN_MARK_BYTE(c, 1.0); //grad = 1.0

cinfo->Al = (c) & 15; //grad = 1.0
...
(*block)[natural_order[k]] = 
                (JCOEF) (v << cinfo->Al);
/* block[0] gradient = 8.0 */
    
void jpeg_idct_islow(int * block) {
    ...
    int * inptr = block; // grad = 8.0

    z2 = (int) inptr[0] * quantptr[0]
    /*z2 gradient = 2040.0 can overflow*/
    
    z2 = z2 << 13;
    /* negative z2 triggers error */
}
\end{lstlisting}
         \setlength{\belowcaptionskip}{-10pt}
    \caption{Arithmetic Error in \bf\texttt{djpeg}.}
    \label{lst:implicit_control}
\end{figure}

\begin{table}
\centering
{\small
 \begin{tabular}{ll r r} 
 \toprule
 & & Integer & Memory \\
  Library & Test Program &  Overflow & Corruption \\ 
 \midrule
 libjpeg-9c         & \verb|djpeg|          & 2 & 3 \\
 mupdf-1.14.0       & \verb|mutool show|    & 1 & 0 \\
 binutils-2.30      & \verb|size|           & 0 & 1 \\  
                    & \verb|objdump -xD|    & 0 & 9 \\  
                    & \verb|strip |         & 0 & 6 \\  
 \bottomrule
 \end{tabular}
 }
 \setlength{\belowcaptionskip}{-10pt} 
  \caption{\label{tab:bug_summary}Summary of new bugs found by PGA. In total there are 22 bugs found over 5 programs.}
  \vspace{0pt}
\end{table}

\subsubsection{Information Leak Discovery}



\revise{We provide two case studies using PGA to detect side channel leaks: one example of a memory usage based side channel in \texttt{objdump} and an execution time based side channel in \texttt{cjpeg}. To identify each information leak, we marked the input file headers as sources and relevant program values as sinks, either memory allocation operands or comparison operands in loops.}

\begin{figure}
  \begin{subfigure}[b]{0.49\columnwidth}
    \includegraphics[width=\linewidth]{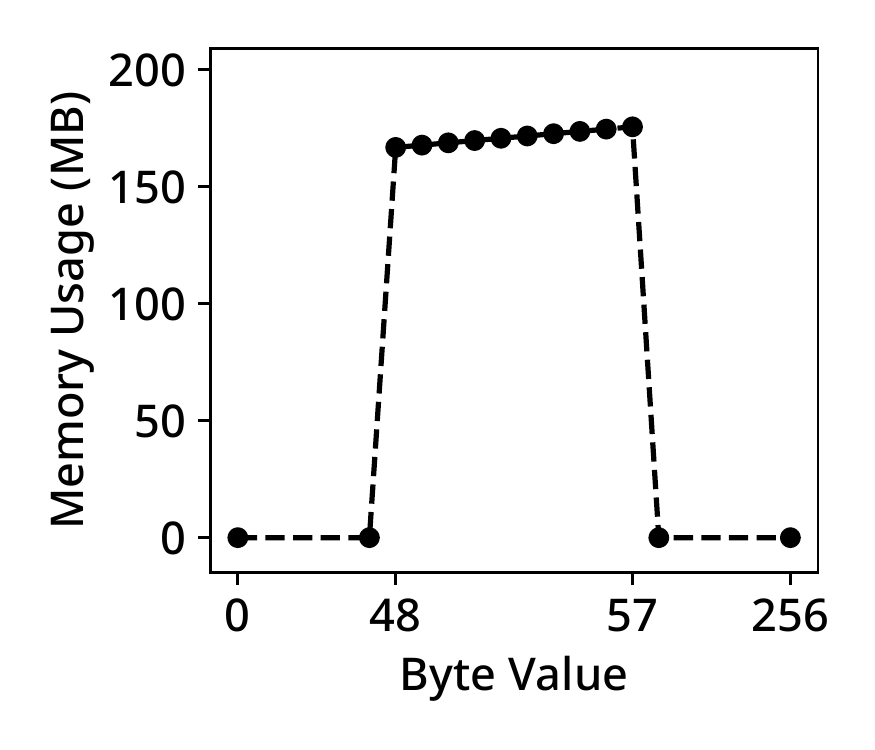}
    \setlength{\abovecaptionskip}{-10pt}
  \caption{\label{fig:memory_side_channel} {\rm \texttt{Objdump} side channel}}
  \end{subfigure}
    \begin{subfigure}[b]{0.49\columnwidth}
    \includegraphics[width=\linewidth]{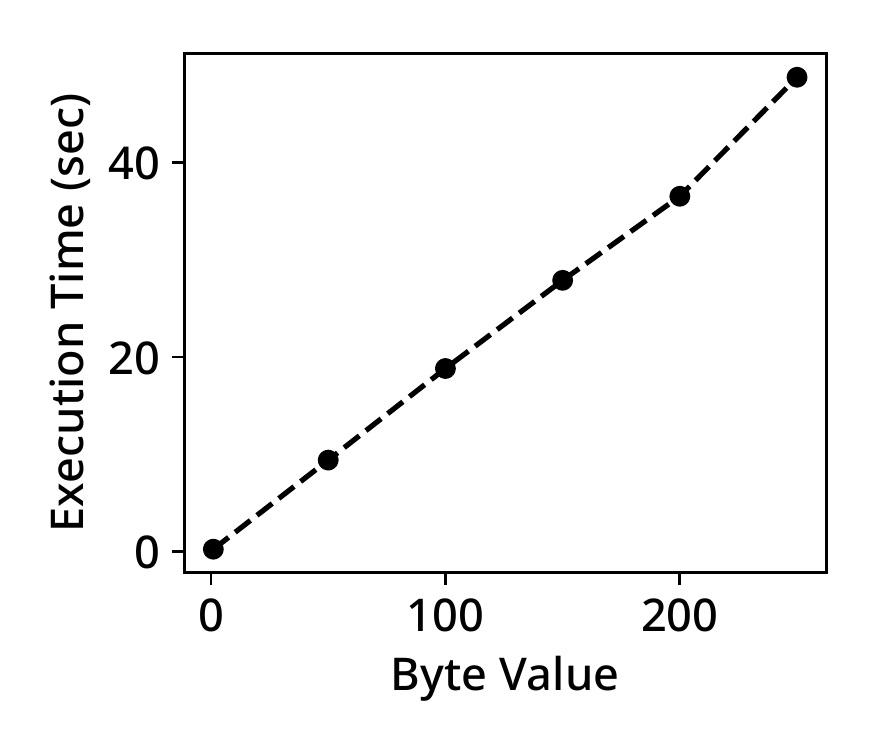}
    \setlength{\abovecaptionskip}{-10pt}
  \caption{\label{fig:time_side_channel} {\rm \texttt{Cjpeg} side channel} }
  \end{subfigure}
  \setlength{\belowcaptionskip}{-10pt}
  \caption{\label{fig:side_channel} Memory and timing side channel leaks.}
\end{figure}

In \texttt{objdump}, we identified a memory based side channel based on a gradient of 1 million to a malloc instruction from the ELF section header for program size.  Figure~\ref{fig:memory_side_channel} shows the effect of incrementing the value from 46 to 59 on the program's total memory usage. The memory consumption is linear in the byte value if the byte is in range from 48 to 57, which can be converted to a valid number '0' to '9' in ASCII. 
Similarly, we identified the timing based side-channel in \texttt{cjpeg} by a gradient from the height field in the jpeg header to the operand of a while loop condition. Figure~\ref{fig:time_side_channel} shows the height information leak in program execution time. 



Prior side channel attacks have demonstrated that these types of leaks can be exploited to learn sensitive information about a user ~\cite{chen2010side, jana2012memento}. For example, one can imagine a malicious Android app that uses JPEG dimensions leaked from a browser to determine which websites the device user is visiting.

\vspace{3pt}
\noindent \begin{longfbox}
{\tbf{Result 6:} PGA successfully detects two information leaks from file headers in \texttt{objdump} and \texttt{cjpeg}.}
\end{longfbox}



\section{Discussion}
\label{sec:discussion}
\revise{
In this section we review the implications of our results and discuss of the relative advantages and limitations of PGA as an approach to dataflow analysis.
}

\noindent \revise{\textbf{Advantages of PGA.} The additional information encoded in gradients can greatly improve precision when predicting dataflows between sources and sinks (i.e. reducing the number of false positives), while the magnitude and direction information can be used to prioritize dataflows based on their significance and predicted effect. We see the benefits of the additional information from gradients in the improved performance of PGA relative to DTA in our dataflow accuracy, guided fuzzing, and vulnerability detection and analysis evaluations (Sections \ref{sec:accuracy}, \ref{sec:fuzzing_eval}, \ref{sec:eval:bug_finding}).
}

\noindent \textbf{Limitations of PGA.} 
While our implementation of PGA demonstrably works based on our evaluation, 
How to best sample non-smooth operations when evaluating proximal gradients is an open question. \rv{Our prototype uses a simple fixed sampling strategy, and does not fully implement proximal gradients on some operations, such as loads on pointers with derivatives (Section \ref{sec:methodology:prop_rules}), which sometimes causes errors in the gradient evaluation. The effect of these errors can be seen in our dataflow accuracy evaluation (Section \ref{sec:accuracy}), where \texttt{grsan} has slightly lower recall than \texttt{dfsan}, indicating some gradients erroneously evaluate to 0.}
We believe incorporating more information about specific operations in sampling strategies, as well as tracking valid domains for some operations, will reduce these errors. 

A second limitation of our implementation of PGA is that, like most DTA frameworks, it does not model implicit dataflows, \rv{such as control flow dependencies}. This can be seen in our CVE evaluation (Section \ref{sec:eval:cves}), where two of our tested CVEs cannot be detected by either PGA or DTA. We intend to explore both more accurate methods for evaluating proximal gradients and modeling implicit dataflows in future work.

\section{Related Work}
\label{sec:related}


\noindent \textbf{Dynamic Taint Analysis.} Dynamic Taint Analysis (DTA) tracks data flow from taint sources to taint sinks at runtime. Common applications of DTA include software vulnerability analysis and information leak detection~\cite{newsome2005dynamic, Yin07panorama, dytan, taintdroid, qsym}. DTA typically overestimates the tainted bytes which contributes to a large performance overhead. Therefore, much of the recent work in DTA has focused on developing more efficient systems~\cite{libdft, Bosman2011MinemuTW, MingStraightTaint}. Like DTA, PGA dynamically propagates dataflow information, but it provides more fine-grained information in the form of gradients. Moreover, PGA is more precise than DTA, which reduces overtainting in large programs.

\revise{Some DTA systems use bit level taint tracking to improve precision at the cost of higher overheads~\cite{yan2014soundness, yadegari2014bit}. Although we have not implemented it in our current prototype, gradients can also be propagated over individual bits based on functional Boolean analysis, and we expect it to offer similar tradeoffs in improved accuracy for higher overheads~\cite{o2014analysis}.}

Recently, automatically learning taint rules has been used to reduce the approximation errors in DTA~\cite{taintinduce}. This approach is orthogonal to ours and could also potentially be applied to learn gradient propagation rules.

\noindent \revise{\textbf{Quantitative Information Flow.} Quantitative Information Flow (QIF) measures the potential transmission of information through a program using entropy based measures such as channel capacity and min-entropy~\cite{mccamant2008quantitative, smith2011quantifying, espinoza2013min}. QIF has primarily been used for detecting information leaks and ensuring the integrity of program secrets~\cite{heusser2010quantifying, alvim2014additive, doychev2015cacheaudit}, but has also been proposed as a way of enhancing taint tracking~\cite{newsome2009measuring}. PGA adds a different type of information as discussed in Appendix~\ref{app:additonal_acc_exps}, and does not have the high computational complexity involved in estimating information flows accurately.}

\noindent \textbf{Gradient-guided fuzzing.} Recent fuzzers have used gradient approximations to guide their mutation process. Angora estimates finite differences, an approximation of gradients with many known limitations especially for high-dimensional problems, by executing the program on modified inputs and recording the changes in the outputs~\cite{angora,diffmethods1967}. \texttt{NEUZZ}, MTFuzz and \texttt{Neutaint} train neural networks to predict program branch behavior and use the network's gradients to guide the mutation algorithm~\cite{neuzz, she2020mtfuzz, she2019neutaint}. This incurs less overhead than instrumentation based methods but is also less exact since it operates on an approximate model of the program. By contrast, PGA computes gradients directly over the program's individual instructions and therefore produces precise gradients.

\noindent \textbf{Program Smoothing.} Prior work has explored computing gradients with smooth interpretation of a program via a Gaussian kernel~\cite{chaudhuri2010smooth, robustsmoothing2011} or parametric relaxation of SMT~\cite{ryan2019cln2inv, yao2020learning}. These methods use symbolic reasoning and have not been applied to analysis of real world programs. PGA's approximation methods are more efficient and have been successfully demonstrated on real world programs. 




\section{Conclusion}
\label{sec:conclusion}

In this paper we introduce proximal gradient analysis (PGA), a novel theoretically-grounded approach to dataflow analysis that uses non-smooth calculus techniques to compute gradients over programs. PGA is more precise than dynamic taint tracking and provides more fine grained information about program behavior. We provide a prototype implementation of PGA based on the LLVM framework and show that it outperforms three state-of-the-art DTA systems in accuracy while adding less than 5\% overhead on average. Finally, we show PGA is an effective tool for security analysis, identifying relevant dataflows for 19 different CVEs, discovering 22 bugs, and detecting 2 side-channel leaks in 7 real world programs. We hope that our approach to program analysis will motivate other researchers to explore new techniques exploiting the rich non-smooth analysis literature.

\section*{Acknowledgements} 
\gabe{specify ndseg and nsf more clearly}
We thank our shepherd Lujo Bauer and the anonymous reviewers for their constructive and valuable feedback. The first author is supported by an NDSEG Fellowship, and the second author is supported by an NSF Graduate Fellowship. This work is sponsored in part by NSF grants CNS-18-42456, CNS-18-01426, CNS-16-17670; ONR grant N00014-17-1-2010; 
an ARL Young Investigator (YIP) award; a NSF CAREER award; a Google Faculty Fellowship; and a Capital One Research Grant, as well as European Union Marie Sklodowska-Curie grant agreement 690972 (PROTASIS) and innovation programme under grant agreement No. 786669 (ReAct). Any opinions, findings, conclusions, or recommendations expressed herein are those of the authors, and do not necessarily reflect those of the US Government, European Union, ONR, ARL, NSF, Google, or Capital One.

\bibliographystyle{plain}
\bibliography{main}

\appendix

\section{Additional Accuracy Experiments}
\label{app:additonal_acc_exps}

\begin{table}[b]
\small{
\begin{tabular}{p{1.3cm} p{0.7cm}p{0.7cm}p{0.7cm}p{0.7cm}p{0.7cm}p{0.7cm}}
\toprule
& \multicolumn{2}{c}{{Opt 0}} &  \multicolumn{2}{c}{{Opt 1}} & \multicolumn{2}{c}{{Opt 2}} \\
&\texttt{dfsan}&\texttt{grsan}&\texttt{dfsan}&\texttt{grsan}&\texttt{dfsan}&\texttt{grsan}\\
\midrule
minigzip&0.39&0.57&0.42&0.51&0.39&0.45\\
djpeg&0.36&0.69&0.29&0.65&0.31&0.63\\
mutool&0.62&0.64&0.56&0.66&0.52&0.62\\
xmllint&0.76&0.92&0.73&0.88&0.74&0.81\\
objdump&0.52&0.71&0.48&0.67&0.47&0.68\\
strip&0.33&0.63&0.31&0.60&0.31&0.61\\
size&0.53&0.74&0.52&0.68&0.51&0.69\\
\bottomrule
\end{tabular}
}
         \setlength{\belowcaptionskip}{-10pt}
 \caption{\label{tab:optimization_comparison} Effect compiler optimization levels on dataflow F1 accuracy. The table shows \texttt{grsan} has significantly higher F1 accuracy than \texttt{dfsan} for all three measured optimization levels ($>18\%$ average)}
\end{table}

We describe the additional accuracy evaluations summarized in Section~\ref{sec:accuracy} here. Specifically, the ablation of gradient information, the effects of compiler optimization, PGA vs. \texttt{Neutaint} in coarse grained dataflow prediction, analysis of 0 gradients, and a comparison with QIF.

\parheader{Gradient Ablation.}
\label{app:binary_grad_ablation}
We measure the effect of the gradient information on determining accurate dataflows by performing an ablation with binary valued gradients. The ablation uses the same proximal gradient propagation rules, but rounds all gradients to 0 or 1. In effect, this converts PGA into DTA with PGA propagation. 

Results of the comparison are shown in Table~\ref{tab:precision_comp}. PGA with floating point gradient information performs significantly better than PGA with binary gradients for every program.
These results indicate that precise gradients are key to the performance gains achieved by PGA because they compose accurately over multiple operations. 

\noindent \textbf{Compiler Optimization.} 
We evaluate the impact of compiler optimization levels on dataflow accuracy at 3 optimization levels: \texttt{-O0}, \texttt{-O1}, and \texttt{-O2}. Table~\ref{tab:optimization_comparison} summarizes the effects of 3 compiler optimization levels on dataflow F1 accuracy. Increasing the compiler optimization levels reduces the accuracy of both PGA and DTA by a small amount (<3.6\%) for both \texttt{-O1} and \texttt{-O2}. On average, PGA is at least 18\% more accurate than DTA for all three tested optimization levels.


\noindent\textbf{\texttt{Neutaint} hotbyte evaluation.} \texttt{Neutaint}'s neural network based approach does not perform well in fine grained dataflow prediction, but is better suited to identifying hot bytes (input bytes that are most influential to program behavior). We therefore perform the hot byte evaluation described in ~\cite{she2019neutaint} on PGA. 
Our results are summarized in Table~\ref{tab:agghotbyte}. On average, PGA predicts hotbytes with 43.75\% accuracy, while \texttt{Neutaint} predicts hotbytes with 64.25\% accuracy. We see \texttt{Neutaint} as a complementary method to PGA, where PGA is better suited to fine grained dataflow prediction 
and both methods could be used together in program analysis.

\begin{table}
\centering
\small{
\begin{tabular}{l r r}
\toprule
Program  & \texttt{Neutaint}    & PGA   \\
\midrule
\texttt{mutool}   & 73\%   & 99\%   \\
\texttt{xmllint }            & 76\%   & 1\%   \\
\texttt{djpeg}         & 37\%  & 33\%  \\
\texttt{miniunz}       & 71\%    & 42\%    \\
\bottomrule
\end{tabular}
}
\vspace{0pt}
 \caption{\label{tab:agghotbyte} Neutaint Hotbyte Evaluation results. On average, \texttt{Neutaint} predicts hot bytes with 64.25\% accuracy and PGA with 43.75\% accuracy. We believe \texttt{Neutaint} outperforms PGA because it makes a prediction based on many program inputs, whereas PGA makes a prediction based on a single input. Note our results different from the original \texttt{Neutaint} paper ~\cite{she2019neutaint} due to different initializations and environments for training the neural network.  }
 \vspace{-10pt}
 \end{table}

\noindent \textbf{Zero gradient analysis.} 
PGA is able to avoid overtainting when it computes a zero gradient on an instruction DTA would mark as tainted. Therefore we investigate the distribution of zero gradients across programs and instruction types to determine where and how PGA is more precise than DTA. For each program and each type of instruction, we count how many times the instruction had zero gradient in the execution traces from the accuracy evaluation.  Table~\ref{tab:overall_zero_grad_insts} shows the results of this analysis for each instruction and program. 



\noindent \textbf{QIF Comparison.} 
We compare PGA with the latest version of a publicly available QIF tool \texttt{Flowcheck}~\cite{mccamant2008quantitative}. 
We perform a similar experiment to Section~\ref{sec:accuracy}, but
since \texttt{Flowcheck} does not 
byte-level granularity, we compute accuracy by aggregating flows over all bytes so that PGA is not unfairly advantaged. 
We outperform \texttt{Flowcheck} in terms of F1 accuracy by 22\% on average on all of the evaluated programs as summarized in Table~\ref{tab:qif_comp}. 

\begin{table}
\small{
\begin{tabular}{l  | ccc | ccc }
\toprule


\multicolumn{1}{c}{ }  & \multicolumn{3}{c}{{\bf Flowcheck}}  & \multicolumn{3}{c}{{\bf PGA}} \\
&Prec.&Rec.&F1&Prec.&Rec.&F1\\
\midrule
minigzip&   0.44&   0.62&   0.52         & 0.7&   0.62&   0.66     \\
djpeg&   0.44&   0.95&   0.6      &    0.62&   0.87&   0.73    \\
mutool&   0.69&   0.77&   0.73 &  0.89&   0.64&   0.74      \\
xmllint&   0.55&   0.08&   0.14  &     0.99&   0.95&   0.97  \\
objdump&   0.71&   0.62&   0.66 &       0.75&   0.79&   0.77    \\
strip&   0.49&   0.65&   0.56           & 0.66&   0.88&   0.75 \\
size&   0.74&   0.64&   0.69     &     0.74&   0.93&   0.82  \\
\bottomrule
\end{tabular}
}
 \caption{\label{tab:qif_comp}QIF accuracy comparison results for PGA and \texttt{Flowcheck}. PGA outperforms \texttt{Flowcheck} by 22\% on average in terms of F1 accuracy.}
\end{table}

\begin{table}
\centering
{\small
  \begin{tabular}{p{0.9cm}p{0.7cm}p{0.8cm}} 
 \toprule
 \multicolumn{3}{c}{Program Summary}\\
 \multicolumn{3}{c}{Over all Instructions}\\
 \toprule 
   Program & Instrs  & \%Zeros  \\ 
 \midrule
minigzip & 3012 & 28.2  \\
djpeg & 703 & 38.7  \\
mutool & 401 & 40.4  \\
xmllint & 430 & 39.5  \\
objdump & 1070 & 39.0  \\
strip & 3089 & 41.0  \\
size & 659 & 19.3  \\
 \bottomrule
 \end{tabular} \hspace{5pt}
}
 {\small
   \begin{tabular}{p{0.7cm}p{0.7cm}p{0.7cm}} 
 \toprule
  \multicolumn{3}{c}{Instruction Summary}\\
  \multicolumn{3}{c}{Across all Programs}\\
  \toprule 
 Instr. & Total & \%Zeros \\ 
 \midrule
 And & 6756 & 30.2 \\
URem & 214 & 29.0 \\
Sub & 1214 & 21.0 \\
Mul & 875 & 15.9 \\
LShr & 2377 & 14.4 \\
AShr & 149 & 6.0 \\
Add & 895 & 5.7 \\
 \bottomrule
 \end{tabular} 
 }
         \setlength{\belowcaptionskip}{-10pt}
  \caption{\label{tab:overall_zero_grad_insts}Analysis of operations from execution traces where gradient drops to 0, aggregated for each program and for each type of instruction across all programs. Outputs of these operations will have 0 gradient but still be marked as tainted by DTA.}
\end{table}

\begin{table}
\centering
\small{
\begin{tabular}{l r r r r}
\toprule
            & \texttt{libdft}   & \texttt{dfsan}    & \texttt{grsan}    & \texttt{grsan} rel. \\
Program     & Overhead & Overhead & Overhead & to \texttt{dfsan}\\
\midrule
minigzip    & 2,379.5\%    & 54.7\%   & 61.5\%   &  4.4\% \\
djpeg       & -            & 70.5\%   & 73.7\%   &  1.9\% \\
mupdf       & 853.5\%      & 198.4\%  & 262.1\%  &  21.5\%\\
xmllint     & 231.4\%      & 5.5\%    & 0.0\%    & -5.2\%\\
size        & 152.5\%      & 101.1\%  & 107.1\%  &  3.0\%\\
objdump     & 180.0\%      & 133.2\%  & 131.2\%  & -0.9\%\\
strip       & 142.5\%      &  12.0\%  & 11.4\%   & -2.2\%\\
\bottomrule
\end{tabular}
}
\vspace{0pt}
 \caption{\label{tab:performance} Program runtime overhead measurements averaged over five runs for a single taint/gradient source. \texttt{Libdft} overhead is measured relative to running a program only with PIN. \texttt{Dfsan} and \texttt{grsan} are measured relative to uninstrumented programs. 
 After 6 hours, \texttt{libdft} execution timed out on \texttt{djpeg}. 
 }
 \vspace{-10pt}
 \end{table}

\section{Runtime and Memory Overhead Evaluation}
\label{app:overhead_evaluation}

\parheader{Program Overhead.} \rv{We evaluate the overhead introduced by our implementation of PGA in runtime and memory and compare it to \texttt{dfsan} for a single taint/gradient source.} To measure overhead, we execute each program while recording runtime and memory usage. For runtime we perform 5,000 executions for each measurement. We perform each measurement 5 times and average the measured runtime and memory usage. 

Tables \ref{tab:performance} and \ref{tab:memory} detail the runtime and memory overhead per program in our evaluation.
In the worst case PGA has 21.7\% greater overhead in runtime and 21.5\% in memory relative to DTA, but on average only adds 3.21 \% relative overhead in runtime and 1.48\% in memory. We also provide overhead measurements for \texttt{libdft}, although it adds significantly more overhead due to the binary instrumentation.


 \begin{table}
\centering
\small{
\begin{tabular}{l r r r}
\toprule
        & \texttt{dfsan}    & \texttt{grsan}    & \texttt{grsan} rel. \\
Program & Overhead & Overhead & to \texttt{dfsan}\\
\midrule
minigzip    & 183.7\%   & 245.3\%   &  21.7\% \\
djpeg             & 276.4\%   & 291.9\%   &  4.1\% \\
mupdf         & 112.4\%  & 124.7\%  &  5.8\%\\
xmllint       & 346.6\%    & 258.5\%    & -19.7\%\\
size           & 373.3\%  & 392.4\%  &  4.0\%\\
objdump        & 345.6\%  & 323.5\%  & -5.0\%\\
strip           &  344.5\%  & 342.1\%   & -0.5\%\\
\bottomrule
\end{tabular}
}
\vspace{0pt}
 \caption{\label{tab:memory} \rrevise{Memory overhead for each program averaged over five runs relative to uninstrumented programs for a single taint/gradient source. \texttt{Grsan} may increase or decrease overhead because gradients require more memory to store, but may use less overall memory due to increased precision. On average, \texttt{grsan} adds 1.48\% additional overhead relative to \texttt{dfsan}. }}
 \vspace{-10pt}
 \end{table}

\section{Evaluation on Current Fuzzers}
\label{app:fuzzing_current_fuzzers}

We also evaluate if the gradient information from PGA can improve the performance of state-of-the-art fuzzers such as \texttt{NEUZZ} and \texttt{VUzzer}. We use \texttt{NEUZZ} as a basis because it has higher edge coverage as seen in Table~\ref{tab:fuzzerimprovement} and already incorporates gradients from a neural network in its mutation strategy. We modify \texttt{NEUZZ} so that it uses the PGA gradients to guide its mutation strategy. We run \texttt{grsan} on its inputs and send the resulting gradients to the \texttt{NEUZZ} backend. \rv{Note that \texttt{NEUZZ} is designed to operate on gradients, so we did not modify it to also use DTA. We provide a controlled comparison of PGA vs. DTA for guided fuzzing in Evaluation 5.2.3.}

We compare the additional edge coverage achieved by the fuzzers over a 24hr run. Since we use some programs with different file formats from the original \texttt{NEUZZ} benchmark, we use a new seed corpus generated by running AFL on each program for 1 hour. We perform this experiment using cloud hosted virtual machines.
Table~\ref{tab:fuzzerimprovement} summarizes the modified \texttt{PGA+NEUZZ} against baseline  \texttt{NEUZZ} and \texttt{VUzzer}.  
On average, \texttt{PGA+NEUZZ} improves new edge coverage by 12.9\% over baseline \texttt{NEUZZ}. We hypothesize this improvement is because the gradients produced with PGA are more precise than the neural-network based gradients used by by \texttt{NEUZZ}. 
The very similar results in edge coverage on \texttt{minigzip} are caused by the CRC check in minigzip, which causes the program to exit early on most new inputs. 
\texttt{VUzzer} crashes on \texttt{minigzip} due to an error in its taint tracking and achieves a low edge coverage for \texttt{djpeg} because of the high overhead of PIN's dynamic binary instrumentation for taint tracking.  Note that our results are slightly different from the original \texttt{NEUZZ} and \texttt{VUzzer} results due to different initial seed corpuses, program versions, and test environments. 

\begin{table}
\centering
\small{
\begin{tabular}{l r r r r}
\toprule
 \multicolumn{5}{c}{\textbf{Edge Coverage after 24hrs}}\\
Program  & \texttt{VUzzer}   & \texttt{NEUZZ}    & \texttt{PGA +} & \texttt{PGA+NEUZZ} \\
         &  &  & \texttt{NEUZZ} & rel. to \texttt{NEUZZ}\\
\midrule
minigzip    & -     & 87   & 94   &  8.1\% \\
djpeg       & 7            & 645   & 686   &  6.4\% \\
mupdf       & 156      & 376 & 430  &  14.4\%\\
xmllint     & 282     & 957    & 1079   & 12.8\%\\
size        & 474     & 1580 & 2064  &  30.6\%\\
objdump     & 247      &1813  & 2014 & 11.1\%\\
strip       & 1337      & 3394  & 3637   &7.2\%\\
\bottomrule
\end{tabular}
}
\vspace{0pt}
 \caption{\label{tab:fuzzerimprovement} \rrevise{New edge coverage for each program over 24 hours by three different fuzzers.
\texttt{VUzzer} encounters an error in its taint tracking on minigzip and crashes.
 Overall, \texttt{PGA+NEUZZ} improves \texttt{NEUZZ} edge coverage on average by 12.9\%. Note that our results are slightly different from the original \texttt{NEUZZ} and \texttt{VUzzer} results due to differences in test environments, input corpuses, and program versions.}}
 \vspace{-10pt}
 \end{table}

\end{document}